\documentclass[12pt]{iopart}
\usepackage{bm}
\usepackage{iopams}
\usepackage{epsf,graphicx}

\begin{document}
\title[Cosine and Sine Operators Related with Orthogonal Polynomial Sets]{Cosine and Sine Operators Related with Orthogonal Polynomial Sets on the intervall [-1,1]}
\author{Thomas Appl and Diethard H Schiller} 
\address{Fachbereich Physik, Universit\"{a}t Siegen, D-57068 Siegen, Germany}
\ead{schiller@physik.uni-siegen.de}

\begin{abstract}
The quantization of phase is still an open problem. In the approach of Susskind and Glogower so called cosine and sine operators play a fundamental r\^{o}le. 
Their eigenstates in the Fock representation are related with the Chebyshev polynomials of the second kind. Here we introduce more general cosine and sine operators whose eigenfunctions in the Fock basis are related in a similar way with arbitrary orthogonal polynomial sets on the intervall $\mbox{[-1,1]}$. 
To each polynomial set defined in terms of a weight function there corresponds a pair of cosine and sine operators. Depending on the symmetry of the weight function
we distinguish generalized or extended operators.
Their eigenstates are used to define cosine and sine representations and probability distributions. We consider also the inverse arccosine and arcsine operators and use their eigenstates to define cosine-phase and sine-phase distributions, respectively. Specific, numerical and graphical results are given for the classical orthogonal polynomials and for particular Fock and coherent states.
\end{abstract}

\submitto{\JPA}

\maketitle

\section{Introduction} \label{S1}

The amplitude and phase of a harmonic oscillator are simple and well understood notions in classical physics. The corresponding canonical pair of variables is given by the action $I$ and angle $\phi$ satisfying the classical Poisson bracket $\{I,\phi \}=1$. Their quantization \cite{di, lo, jo} turned out to be a very difficult problem for which a satisfactory solution is still lacking. 
Associating - as suggested by the correspondence principle - the number operator $\hat N$ with the action variable, it turns out to be impossible to define a canonically conjugate Hermitian phase operator $\it \hat \Phi$ such that 
the commutation relation ${[\hat N, \it \hat \Phi]}_{-}=\rmi \hat 1$ holds. 
As a consequence, many different and more or less satisfactory approaches to the quantum phase problem have been put forward. For a review of the various formalisms  we refer to the literature \cite{cn}-\cite{plp}.

One of the most `complete and reasonably consistent' theories is supplied by the Susskind--Glogower formalism which has `become something of a standard formalism for discussing phase problems' (citation from \cite{ly}). 
The main idea is to consider Hermitian cosine and sine operators instead of the questionable phase operator itself. The idea was suggested by Louisell \cite{lu}, implemented independently by Susskind and Glogower \cite{sg} and generalized by Lerner \cite{le, lhw}.

Here we present a generalization of the Susskind--Glogower formalism from a different point of view.
The Hermitian cosine $\hat C_{\rm SG}$ and sine $\hat S_{\rm SG}$ operators introduced by Susskind and Glogower in terms of the exponential operators, $\hat E_{\rm SG}$ and $\hat E_{\rm SG}^{\dagger}$, are given in the Fock basis\footnote{The Fock basis is formed by the eigenstates $\vert n \rangle$, $n=0,1,2,\ldots$, of the number operator $\hat N$.
The states obey the relations of orthonormality 
$\langle n \vert m \rangle = \delta_{nm}$ and completeness 
$\sum_{n=0}^{\infty} \vert n \rangle \langle n \vert = \hat 1$.} 
by
\begin{eqnarray}
\hat C_{\rm SG} \, &=& \, \frac{1}{2} \, (\hat E_{\rm SG}+\hat E_{\rm SG}^{\dagger})\ = \,
\frac{1}{2} \, \sum_{n=0}^{\infty} \, (\vert n \rangle \langle n+1 \vert + \vert n+1 \rangle \langle n \vert ) ,
\label{G10a} \\
\hat S_{\rm SG} \, &=& \, \frac{1}{2 \, \rmi} \, (\hat E_{\rm SG}-\hat E_{\rm SG}^{\dagger}) =
\frac{1}{2 \, \rmi}  \sum_{n=0}^{\infty} \, (\vert n \rangle \langle n+1 \vert - \vert n+1 \rangle \langle n \vert ) .
\label{G10b}
\end{eqnarray}
The corresponding eigenvalue equations 
$\hat C_{\rm SG} \vert c \rangle_{\rm SG} = c  \vert c \rangle_{\rm SG}$
and $\hat S_{\rm SG} \vert s \rangle_{\rm SG} = s  \vert s \rangle_{\rm SG}$
are solved, respectively, by the eigenstates \cite{plp}
\begin{equation}
\fl
\vert c \rangle_{\rm SG} = \sqrt{\frac{2}{\pi}} \, \sqrt[4]{1-c^2} \, \sum_{n=0}^{\infty} \, U_{n}(c) \, \vert n \rangle ,  \qquad \quad
\vert s \rangle_{\rm SG} \, = \, \sqrt{\frac{2}{\pi}} \, \sqrt[4]{1-s^2} \, \sum_{n=0}^{\infty} \, \rmi^n \, U_{n}(s) \, \vert n \rangle ,
\label{G18} 
\end{equation}
where $U_{n}(x)$ for $x=c$ or $s$ are the Chebyshev polynomials of the second kind. 
According to (\ref{G18}) the $n$th Fock component of the eigenstates, $\langle n \vert x \rangle_{\rm SG}$, is proportional to $U_{n}(x)$, with an additional factor $\rmi^n$ for the sine states. 
It is in this sense that we say the Susskind--Glogower cosine and sine operators are related with the Chebyshev polynomials of the second kind. 
In this paper we extend this relationship to arbitrary sets of orthogonal polynomials on the intervall $\mbox{[-1,1]}$. The restriction to this intervall follows from the envisaged interpretation of $x$ as a cosine $c$ or sine $s$ variable.

In section \ref{S2} we summarize the relevant properties of the orthogonal polynomial sets on the intervall $\mbox{[-1,1]}$. In section \ref{S3} we introduce the cosine and sine operators related with such polynomial sets and discuss their properties. The corresponding cosine and sine probability distributions are considered in section \ref{S4}. 
The inverse arccosine and arcsine operators and the corresponding cosine-phase and sine-phase 
probability distributions are studied in section \ref{S5}. Numerical results and graphical representations are given in sections \ref{S3} to \ref{S5} for particular classical orthogonal polynomial sets and for chosen Fock and coherent states. We conclude with a summary in section \ref{S6}.

\section{Orthogonal polynomial sets -- a reminder} \label{S2}

In this section we summarize the relevant properties of the orthogonal polynomial sets  \cite{as, sz, chi} on the intervall $\mbox{[-1,1]}$ and give explicit expressions for the corresponding classical orthogonal polynomials used as examples.

Let $w(x)$ be a real-valued non-negative weight function on the intervall $x \in \mbox{[-1,1]}$ with the property that all its moments exist and are finite
\begin{equation}
\mu_{n} = \int_{-1}^{+1} \rmd x \ w(x) \, x^{n}\, < \infty  ,  \qquad \qquad n=0,1,\ldots 
\label{G37}
\end{equation}
Let $P_{n}(x)$ denote a real-valued polynomial of degree $n$. The set or system of polynomials $\{P_{n}(x)\}$, $n=0,1,2, \ldots $ is said to be orthogonal with respect to $w(x)$, if 
\begin{equation}
\int_{-1}^{+1} \rmd x \, w(x) \, P_{n}(x) \, P_{m}(x) \, = \, d_{n} \, \delta_{n,m}, \label{G36}
\end{equation}
where $d_{n}>0$ is a finite normalization constant and $\delta_{n,m}$ is the Kronecker symbol. 
The weight function $w(x)$ determines the set $\{P_{n}(x)\}$ up to a constant factor in each polynomial $P_{n}(x)$. Different sets of orthogonal polynomials are obtained by choosing different weight functions. Each set can be obtained from the fundamental set of non-negative powers $\{1,x,x^2, \ldots  \}$ by the Gram--Schmidt orthogonalization method.

The polynomials of a given set satisfy the completeness relation
\begin{equation}
\sum_{n=0}^{\infty} \, \frac{1}{d_{n}} \, P_{n}(x)\, P_{n}(x') \, = \, 
\frac{\delta(x-x')}{w(x')}  \label{G40}
\end{equation}
and the recurrence relation connecting three consecutive polynomials \cite{as}
\begin{equation}
P_{n+1}(x) = (\alpha_{n} + x \, \beta_{n})\, P_{n}(x) - \gamma_{n}\, P_{n-1}(x) ,  \qquad n=0,1,2, \ldots \label{G44}
\end{equation}
with $\gamma_{0} = 0$ by definition. The recurrence coefficients are given by
\begin{equation}
\beta_{n}  = \frac{a_{n+1}}{a_{n}} \, , \qquad 
\alpha_{n} = \beta_{n} \left( \frac{b_{n+1}}{a_{n+1}} - \frac{b_{n}}{a_{n}} \right), \qquad 
\gamma_{n} = \frac{\beta_{n} d_{n}}{\beta_{n-1} d_{n-1}} \, , \label{G50}
\end{equation}
where $a_{n}$ and $b_{n}$ are, respectively, the coefficients of $x^n$ and $x^{n-1}$ in the polynomial 
$P_{n}(x) = a_{n} x^n + b_{n} x^{n-1} +  \cdots $.
Of particular interest are the polynomial sets generated by even weight functions, $w(x)=w(-x)$. In this case all moments of odd order vanish and the polynomials of even (odd) degree $n$ contain only even (odd) powers of $x$, implying $b_{n}=0$ and thus $\alpha_{n}=0$ for all $n$. The polynomials then have a definite parity according to the relation $P_{n}(-x)=(-1)^n \, P_{n}(x)$.

Related with the set of orthogonal polynomials $\{P_{n}(x)\}$ we now introduce the set of functions $\{p_{n}(x)\}$ (which are not polynomials, in general) defined by
\begin{equation}
p_{n}(x) \, \equiv \, 
\sqrt{w(x)} \, \bar P_{n}(x) , \qquad \qquad
\bar P_{n}(x) \equiv P_{n}(x) / \sqrt{d_{n}} \, ,
\label{G52}
\end{equation}
with $\bar P_{n}(x)$ the \emph{orthonormal} polynomials. The functions $p_{n}(x)$ are orthonormal 
\begin{equation}
\int_{-1}^{+1} \rmd x \ p_{n}(x) \, p_{m}(x) \, = \, \delta_{n,m} ,
\label{G56}
\end{equation}
satisfy the completeness relation
\begin{equation}
\sum_{n=0}^{\infty} \, p_{n}(x) \, p_{n}(x') \, = \, \delta(x-x')  \label{G60}
\end{equation}
and the three-term recurrence formula
\begin{equation}
\frac{f_{n}}{2} \, p_{n+1}(x)+ \frac{f_{n-1}}{2} \, p_{n-1}(x) + g_{n} \, p_{n}(x) \, = \, x \, p_{n}(x) . 
\label{G66}
\end{equation}
Here $f_{-1} \equiv 0$ by definition and the other recurrence coefficients are given by
\begin{equation}
\frac{f_{n}}{2}\, = \, \frac{a_{n}}{a_{n+1}} \, \sqrt{\frac{d_{n+1}}{d_{n}}} \, , \qquad \qquad
g_{n} \, = \, \frac{b_{n}}{a_{n}} - \frac{b_{n+1}}{a_{n+1}} \, . 
\label{G70}
\end{equation}
Note that $g_{n}=0$ for the polynomial sets generated by even weight functions. Equation (\ref{G66}) contains only two independent sets of coefficients, $\{f_{n}\}$ and $\{g_{n}\}$, and is actually the well known recurrence relation for the orthonormal polynomials $\bar P_{n}(x)$, since the overall factor $\sqrt{w(x)}$ cancels. 
For the considerations in the next section it is crucial that the coefficients of $p_{n+1}(x)$ and $p_{n-1}(x)$ turn out to be, respectively, $f_{n}$ and $f_{n-1}$ taken from the same set $\{f_{n}\}$.

\begin{table}
\renewcommand{\arraystretch}{2.2}
\caption{\label{T1}Weight function $w(x)$, normalization constant $d_{n}$ and recurrence coefficients $f_{n}$ and $g_{n}$ for the Jacobi polynomials $P_{n}^{(\mu,\nu)}(x)$.}
\noindent
\begin{tabular}{l}
\br
$ P_{n}^{(\mu,\nu)}(x) \qquad \mu > -1, \ \nu > -1$ \\
\mr
$ w(x) \ = \ (1-x)^{\mu} \ (1+x)^{\nu}$ \\
$ \displaystyle \ \ d_{n} \ \ = \ \frac{2^{\, \mu+\nu+1} \Gamma(n+\mu+1)\Gamma(n+\nu+1)}{n!(2n+\mu+\nu+1) \Gamma(n+\mu+\nu+1)}$ \\
$ \displaystyle \ \ f_{n} \ \ = \ \frac{4}{2n+\mu+\nu+2} \ 
\sqrt {\frac{(n+1)(n + \mu + 1)(n + \nu + 1)(n + \mu + \nu + 1)}
{(2n + \mu + \nu + 1)(2n + \mu + \nu + 3)}}$ \\
$ \displaystyle \ \ g_{n} \ \ = \ 
\frac{\nu^2 - \mu^2}{(2n + \mu + \nu)(2n + \mu + \nu + 2)}  \qquad 
\Bigl(g_{0} = \frac{\nu - \mu}{\mu + \nu + 2}\Bigr)$  \\ 
\br
\end{tabular} \\
\end{table}

The class of orthogonal polynomial sets on the intervall $\mbox{[-1,1]}$ is very large. For illustrative purposes we will restrict ourselves to the remarkable sets of the \emph{classical} orthogonal polynomials. 
The most general classical polynomials on $\mbox{[-1,1]}$ are the Jacobi polynomials $P_{n}^{(\mu,\nu)}(x)$ with characteristics given in Table~\ref{T1} for unequal indices ($\mu \ne \nu$) and in Table~\ref{T2} for equal indices ($\mu=\nu$). Table~\ref{T2} contains also the entries for the Gegenbauer ($\lambda = \mu + \frac{1}{2}$), Legendre ($\lambda = \frac{1}{2}$) and Chebyshev polynomials of the first ($\lambda \rightarrow 0$) and second ($\lambda = 1$) kind denoted, respectively, by $C_{n}^{(\lambda)}(x)$, $P_{n}(x)$, $T_{n}(x)$ and $U_{n}(x)$. For real-valued polynomials the parameters $\mu$, $\nu$ and $\lambda$ must be real. The restrictions $\mu > -1$, $\nu > -1$ and $\lambda > - \frac{1}{2}$ follow from the condition of finite moments (\ref{G37}). 
The weight functions and polynomials in Table~\ref{T2} have a definite parity implying $p_{n}(-x)=(-1)^n \, p_{n}(x)$, whereas those in Table~\ref{T1} lead to 
$p_{n}^{(\mu,\nu)}(-x) = (-1)^n \, p_{n}^{(\nu,\mu)}(x)$.

Besides the set of functions $\{p_{n}(x)\}$ we need two other sets related to it by a transformation of the variable $x$. Since $x \in \mbox{[-1,1]}$, we may write $x=\cos \theta_{c}$, with angle $\theta_{c} \in [0,\pi]$ and (positive) Jacobian $\sin \theta_{c}$. The functions defined by 
\begin{equation}
	c_{n}(\theta_{c}) \equiv \sqrt{\sin \theta_{c}}\ p_{n}(\cos \theta_{c}) = 
	\sqrt{\sin \theta_{c} \, w(\cos \theta_{c})}\ \bar P_{n}(\cos \theta_{c})
	\label{G71}
\end{equation}
then satisfy the following orthonormality and completeness relations
\begin{eqnarray}
	\int_{0}^{\pi} \rmd \theta_{c} \ c_{n}(\theta_{c}) \, c_{m}(\theta_{c}) \, = \, \delta_{n,m} , \label{G71a} \\
	\sum_{n=0}^{\infty} \, c_{n}(\theta_{c}) \, c_{n}(\theta_{c}') \, = \, \delta(\theta_{c}-\theta_{c}') . \label{G71b} 	
\end{eqnarray}
Similarly, the transformation $x=\sin \theta_{s}$, with angle $\theta_{s} \in [-\pi /2,\pi /2]$ and (positive) Jacobian $\cos \theta_{s}$, leads for the functions defined by
\begin{equation}
	s_{n}(\theta_{s}) \equiv \sqrt{\cos \theta_{s}}\, p_{n}(\sin \theta_{s})=
	\sqrt{\cos \theta_{s} \, w(\sin \theta_{s})}\ \bar P_{n}(\sin \theta_{s})
	\label{G72}
\end{equation}
to the following orthonormality and completeness relations
\begin{eqnarray}
	\int_{-\pi/2}^{+\pi/2} \rmd \theta_{s} \ s_{n}(\theta_{s}) \, s_{m}(\theta_{s}) \, = \, \delta_{n,m}  , \label{G72a} \\
	\sum_{n=0}^{\infty} \, s_{n}(\theta_{s}) \, s_{n}(\theta_{s}') \, = \, \delta(\theta_{s}-\theta_{s}') . \label{G72b} 
\end{eqnarray}
The functions $c_{n}(\theta_{c})$ and $s_{n}(\theta_{s})$ are related by an angular shift
\begin{equation}
c_{n}(\theta_{c}) = [s_{n}(\theta_{s})]_{\theta_{s}=\frac{\pi}{2} - \theta_{c}} \, ,
\qquad \qquad
s_{n}(\theta_{s}) = [c_{n}(\theta_{c})]_{\theta_{c}=\frac{\pi}{2} - \theta_{s}} \, .
\label{G77}	
\end{equation}
The sets of functions $\{p_{n}(x)\}$, $\{c_{n}(\theta_{c})\}$ and $\{s_{n}(\theta_{s})\}$ provide an orthonormal basis in the Hilbert spaces $\textbf{L}^2(-1,1)$, $\textbf{L}^2(0,\pi)$ and $\textbf{L}^2(-\pi/2,\pi/2)$, respectively.

\begin{table}
\renewcommand{\arraystretch}{2.2}
\caption{\label{T2}Weight function $w(x)$, normalization constant $d_{n}$ and recurrence coefficient $f_{n}$ for the Jacobi $P_{n}^{(\mu,\mu)}(x)$, Gegenbauer $C_{n}^{(\lambda)}(x)$, Legendre $P_{n}(x)$ and Chebyshev polynomials of the first $T_{n}(x)$ and second $U_{n}(x)$ kind.}
\noindent\begin{tabular}{llll}
\br
Polynom. & $w(x)$ & $d_{n}$ & $f_{n}$  \\
\mr
$P_{n}^{(\mu,\mu)}$ & $(1-x^2)^{\mu}$ & $ \displaystyle \frac{2^{2\mu+1} [\Gamma(n+\mu+1)]^2}{n! (2n+2\mu+1) \Gamma(n+2\mu+1)} $ & $ \displaystyle \sqrt{\frac{(n+1)(n+2\mu+1)}{(n+\mu+\frac{1}{2})(n+\mu+\frac{3}{2})}} $ \\
$ C_{n}^{(\lambda)}$   & $ (1-x^2)^{\lambda-\frac{1}{2}}$ & $ \displaystyle \frac{\pi 2^{1-2\lambda} \Gamma(n+2\lambda)}{n! (n + \lambda) [\Gamma(\lambda)]^2}$
& $ \displaystyle \sqrt{\frac{(n+1)(n+2\lambda)}{(n+\lambda)(n+\lambda+1)}} $  \\
$ P_{n}$ & $1$ & $ \displaystyle \frac{2}{2n+1}$ & $ \displaystyle \frac{n+1}{\sqrt{(n+\frac{1}{2})(n+\frac{3}{2})}}$ \\
$ {T_{n}} $ & $ (1-x^2)^{-\frac{1}{2}}$ & $ \displaystyle \frac{\pi}{2} \, \tau_{n} \qquad (\tau_{0}=2, \tau_{n\geq 1}=1)$ & $ \displaystyle \sqrt{\tau_{n}}$ \\
$ U_{n}$ & $ (1-x^2)^{\frac{1}{2}}$ & $ \displaystyle \frac{\pi}{2}$ & $ 1$ \\
\br
\end{tabular} \\
\end{table}

\section{Generalized and extended cosine and sine operators} \label{S3}

In this section we introduce cosine and sine operators related with arbitrary orthogonal polynomial sets on the interval \mbox{[-1,1]}, discuss their eigenstates and consider various operator relations and expectation values.

By analogy with the Susskind--Glogower states (\ref{G18}) we define the states
\begin{equation}
\vert c \rangle \, \equiv \, \sum_{n=0}^{\infty} \, p_{n}(c) \, \vert n \rangle , 
\qquad \quad \ \qquad
\vert s \rangle \, \equiv \, \sum_{n=0}^{\infty} \, \rmi^n \, p_{n}(s) \, \vert n \rangle , \label{G96}
\end{equation}
in terms of the functions (\ref{G52}).
They satisfy the relations of orthogonality
\begin{equation}
\langle c \vert c' \rangle \, = \, \delta(c-c') , \qquad \qquad \qquad
\langle s \vert s' \rangle \, = \, \delta(s-s') , 
\label{G110}
\end{equation}
due to the completeness relation (\ref{G60}), and resolve the identity
\begin{equation}
\int_{-1}^{+1} \rmd c \, \vert c \rangle \langle c \vert \, = \, \hat 1 , 
\qquad \qquad \qquad
\int_{-1}^{+1} \rmd s \, \vert s \rangle \langle s \vert \, = \, \hat 1 , 
\label{G112}
\end{equation}
due to the orthonormality relation (\ref{G56}).
The eigenstates of Susskind and Glogower are recovered for the Chebyshev polynomials of the second kind, $U_{n}(x)$, in Table~\ref{T2}.

We now determine Hermitian operators $\hat C$ (cosine) and $\hat S$ (sine) 
by requiring the eigenvalue equations 
$\hat C \, \vert c \rangle \, = \, c \, \vert c \rangle$ and 
$\hat S \, \vert s \rangle \, = \, s \, \vert s \rangle$ 
to be satisfied on account of the recurrence formula (\ref{G66}). We find
\begin{eqnarray}
\hat C \, &=& \, \sum_{n=0}^{\infty} \, \Bigl\{ \frac{f_{n}}{2} \, \Bigl(\vert n \rangle \langle n+1 \vert + \vert n+1 \rangle \langle n \vert \Bigr) + g_{n} \vert n \rangle \langle n \vert \Bigr\} , \label{G90a} \\
\hat S \, &=& \, \sum_{n=0}^{\infty} \, \Bigl\{ \frac{f_{n}}{2 \, \rmi} \, \Bigl(\vert n \rangle \langle n+1 \vert - \vert n+1 \rangle \langle n \vert \Bigr) + g_{n} \vert n \rangle \langle n \vert \Bigr\} . \label{G90b}
\end{eqnarray}
The operators consist of an (usual) off-diagonal part of lowering $\hat E$ and raising $\hat E^{\dagger}$ operators, and an (unusual) diagonal part $\hat E_{0}$ such that
\begin{equation}
\hat C = \frac{1}{2}(\hat E + \hat E^{\dagger }) + \hat E_{0} \, , \qquad \qquad
\hat S = \frac{1}{2\, \rmi}(\hat E - \hat E^{\dagger }) + \hat E_{0} \, ,
\label{G94}
\end{equation}
\begin{equation}
\fl \hat E = \sum_{n=0}^{\infty} \ f_{n} \ \vert n \rangle \langle n+1 \vert \, , \qquad         
\hat E^{\dagger} = \sum_{n=0}^{\infty} \ f_{n} \ \vert n+1 \rangle \langle n \vert \, , \qquad
\hat E_{0} = \sum_{n=0}^{\infty} \ g_{n} \ \vert n \rangle \langle n \vert \, .
\label{G92}
\end{equation}
These expressions are a direct consequence of the recurrence relation in orthonormal form (\ref{G66}), 
leading to a raising operator that is the Hermitian conjugate of the lowering operator. The diagonal part is absent ($\hat E_{0}=0$) for polynomial sets generated by even weight functions ($g_{n}=0$). 
We call the operators $\hat C$ and $\hat S$ without diagonal terms ($g_{n}=0$) the \emph{generalized}, and those with diagonal terms ($g_{n} \ne 0$) the \emph{extended} cosine and sine operators. 
Similarly, we distinguish generalized and extended exponential operators given by $\hat C \pm {\rm i} \hat S$ and $(\hat C - \hat E_{0}) \pm {\rm i} (\hat S - \hat E_{0})$, respectively.
The generalized operators may be considered as proper generalizations of the Susskind--Glogower operators (corresponding to $f_{n}=1$, $g_{n}=0$), and as instancies of the generalized operators introduced by Lerner \cite{le, lhw}. 
The extended operators, however, define a more general class due to the dependence on two coefficient sequences, $\{f_{n}\}$ and $\{g_{n}\}$. 
The considerations of this paper apply to both types of operators, allowing a unified treatment.

The cosine and sine operators have a simple matrix representation in the Fock basis given by the tridiagonal Jacobi matrices
\begin{equation}
\fl \hat C = \frac{1}{2}
\left( \begin{array}{ccccc}
2g_{0} & f_{0}   & 0        & 0       & \ldots \\
f_{0}  &  2g_{1} & f_{1}    & 0       & \ldots \\
0      & f_{1}   & 2g_{2}   & f_{2}   & \ldots \\
0      & 0       & f_{2}    & 2g_{3}  & \ldots \\
\vdots & \vdots  & \vdots   & \vdots  & \ddots
\end{array} \right) , \quad 
\hat S = \frac{1}{2}
\left( \begin{array}{ccccc}
2g_{0}       & - \rmi f_{0} & 0             & 0             & \ldots \\
+ \rmi f_{0} & 2g_{1}       & - \rmi f_{1}  & 0             & \ldots \\
0            & + \rmi f_{1} & 2g_{2}        & - \rmi f_{2}  & \ldots \\
0            & 0            & + \rmi f_{2}  & 2g_{3}        & \ldots \\
\vdots       & \vdots       & \vdots        & \vdots        & \ddots
\end{array} \right) . 
\label{G130}
\end{equation}
The matrix eigenproblems $\hat C \, \Psi^{(\rm C)}(c) = c \, \Psi^{(\rm C)}(c)$ [$\hat S \, \Psi^{(\rm S)}(s) = s \, \Psi^{(\rm S)}(s)$] lead again to the recurrence formula (\ref{G66}) for the components $p_{n}(c)$ [$\, {\rm i}^n \,  p_{n}(s)$] of the eigenvector-columns $\Psi^{(\rm C)}(c)$ [$\Psi^{(\rm S)}(s)$]. 
Hence, cosine and sine operators can be related also with tridiagonal Jacobi matrices of the form (\ref{G130}) having as spectrum the interval \mbox{[-1,1]}. Generalized (extended) operators then correspond to matrices with vanishing (nonvanishing) entries on the main diagonal.

We now consider various operator relations and quote them for the extended operators. The more compact relations for the generalized operators can be obtained by setting $g_{n}=0$ for all $n$. 
We start with the operators squared
\begin{eqnarray}
\hat C^2 \, &=& \, \frac{1}{4} \sum_{n=0}^{\infty} (f_{n}^2+f_{n-1}^2+4g_{n}^2) \, 
\vert n \rangle \langle n \vert \nonumber \\
&& + \frac{1}{2} \sum_{n=0}^{\infty} f_{n} (g_{n}+g_{n+1}) \,  
(\vert n \rangle\langle n+1 \vert + \vert n+1 \rangle\langle n \vert) \nonumber \\
&& + \frac{1}{4} \sum_{n=0}^{\infty} f_{n} f_{n+1} \, 
( \vert n \rangle \langle n+2 \vert + \vert n+2 \rangle \langle n \vert ) , \label{G150} \\
\hat S^2 \, &=& \, \frac{1}{4} \sum_{n=0}^{\infty} (f_{n}^2+f_{n-1}^2+4g_{n}^2) \, 
\vert n \rangle \langle n \vert \nonumber \\
&& - \frac{\rmi}{2} \sum_{n=0}^{\infty} f_{n} (g_{n}+g_{n+1}) \, 
(\vert n \rangle\langle n+1 \vert - \vert n+1 \rangle\langle n \vert) \nonumber \\
&& - \frac{1}{4} \sum_{n=0}^{\infty} f_{n} f_{n+1} \, 
( \vert n \rangle\langle n+2 \vert + \vert n+2 \rangle\langle n \vert ) . \label{G151} 
\end{eqnarray}
It follows that the classical relation 
$\hat C^2 + \hat S^2 = \hat 1$ 
cannot be satisfied for any choice of the parameters $f_{n}$ and $g_{n}$. In fact, the only choice \cite{lhw} which would do this 
($f_{2n}=\sqrt 2$, $f_{2n+1}=0$, $g_{n}=0$)
is incompatible with the recurrence formula (\ref{G66}).

The commutator and anticommutator are given by
\begin{eqnarray}
{[\hat C, \hat S]}_{-} \, &=& \, \frac{\rmi}{2} \sum_{n=0}^{\infty} (f_{n}^2 - f_{n-1}^2) \, \vert n \rangle\langle n \vert \nonumber \\
&& - \frac{1}{2} \sum_{n=0}^{\infty} f_{n} (g_{n}-g_{n+1}) \, (\vert n \rangle\langle n+1 \vert - \vert n+1 \rangle\langle n \vert) \nonumber \\
&& - \frac{\rmi}{2} \sum_{n=0}^{\infty} f_{n} (g_{n}-g_{n+1}) \,  
(\vert n \rangle\langle n+1 \vert + \vert n+1 \rangle\langle n \vert) , 
\label{G160} \\
{[\hat C, \hat S]}_{+} \, &=& \, 2 \sum_{n=0}^{\infty} g_{n}^2 \, \vert n \rangle \langle n \vert - \frac{\rmi}{2} \sum_{n=0}^{\infty} f_{n} f_{n+1} \, (\vert n \rangle\langle n+2 \vert - \vert n+2 \rangle\langle n \vert)  \nonumber \\
&& + \frac{1}{2} \sum_{n=0}^{\infty} f_{n} (g_{n}+g_{n+1}) \, (\vert n \rangle\langle n+1 \vert + \vert n+1 \rangle\langle n \vert) \nonumber \\
&& - \frac{\rmi}{2} \sum_{n=0}^{\infty} f_{n} (g_{n}+g_{n+1}) \, (\vert n \rangle\langle n+1 \vert - \vert n+1 \rangle\langle n \vert) .
\label{G161}
\end{eqnarray}
Again, it is not possible to obtain the classical relation ${[\hat C, \hat S]}_{-}=0$. The commutator is not even diagonal unless $g_{n}=0$.

Next we consider some (anti)commutators involving the number operator $\hat N$:
\begin{eqnarray}
\ {[\hat N, \hat C]}_{-} &=& - \rmi \, (\hat S - \hat E_{0}) , 
\label{G170a} \\
\ {[\hat N, \hat S]}_{-} &=& + \rmi \, (\hat C - \hat E_{0}) ,
\label{G170b} \\
\ {[\hat N, \hat C]}_{+} &=& \sum_{n=0}^{\infty} \Bigl[ (2n+1) \frac{f_{n}}{2} \, 
(\vert n \rangle\langle n+1 \vert + \vert n+1 \rangle\langle n \vert)
+ 2n g_{n} \, \vert n \rangle \langle n \vert \Bigr] , \label{G170c} \\
\ {[\hat N, \hat S]}_{+} &=& \sum_{n=0}^{\infty} \Bigl[ (2n+1) \frac{f_{n}}{2\, \rmi} \, (\vert n \rangle\langle n+1 \vert - \vert n+1 \rangle\langle n \vert)
+ 2n g_{n} \, \vert n \rangle \langle n \vert \Bigr] . \label{G170d} 
\end{eqnarray}
The commutators correspond to the classical relations only if $\hat E_{0}=0$. The mode equations following from (\ref{G170a}) and (\ref{G170b}) with $\hat N$ as `Hamiltonian' are given by
\begin{equation}
\ddot{\hat C} + \, \hat C = \hat E_{0}\, , \ \qquad  \qquad \qquad \qquad
\ddot{\hat S} + \, \hat S = \hat E_{0}\, .
\end{equation}
These are the equations of motion for a harmonic oscillator with equilibrium position given by $\hat E_{0}$
for $g_{n}\neq 0$ and by the origin for $g_{n}=0$. This gives a nice physical and distinctive interpretation of the generalized and extended trigonometric operators as normal mode and displaced mode coordinates, respectively. This interpretation is also supported by the fact that, due to ${[\hat N, \hat E_{0}]}_{-}=0$, the operators $\hat C - \hat E_{0}$ and $\hat S - \hat E_{0}$ obey the usual commutation relations with $\hat N$ and hence the normal mode equations.
The operators $\hat C$ and $\hat S$ and their eigenstates are in quadrature, i.e. related by a rotation through an angle $\pi/2$ with generator $\hat N$:
\begin{eqnarray}
\hat C \, &=& \ \rme^{- \rmi \frac{\pi}{2} \hat N} \ \hat S \ \rme^{+ \rmi \frac{\pi}{2} \hat N} , \qquad \qquad \quad
\hat S \ \, = \ \rme^{+ \rmi \frac{\pi}{2} \hat N} \ \hat C \ \rme^{- \rmi \frac{\pi}{2} \hat N} , \label{G25} \\
\vert c \rangle \, &=& \, [\, \rme^{- \rmi \frac{\pi}{2} \hat N} \,  \vert s \rangle \, ]_{s=c} \, , \qquad \qquad \ \quad
\vert s \rangle \, = \, [\, \rme^{+ \rmi \frac{\pi}{2} \hat N} \,  \vert c \rangle \, ]_{c=s} \, .  \label{G28}
\end{eqnarray}
This explains nicely the factor $\rm i^{n}$ in the $n$th Fock component of the sine states. Also, being a unitary transformation it relates the sine state with eigenvalue $s$ to the cosine state with the same eigenvalue $c=s$ instead of $c=\sqrt{1-s^2}$. This is one reason why the operators fail to reproduce certain trigonometric relations.

We now come to the expectation values for a system described by the density operator $\hat \rho$.
The expectation value of an operator $\hat A$ is given by the trace
$\langle \hat A \rangle_{\rho} = {\rm Sp}(\hat \rho \hat A)$,  
its variance squared by 
$\sigma_{AA}^{\rho} = \langle \hat A^2 \rangle_{\rho} - \langle \hat A \rangle_{\rho}^2$ 
and the correlation of two operators $\hat A$ and $\hat B$ by 
$\sigma_{AB}^{\rho} = \frac{1}{2} \langle {[\hat A,\hat B]}_{+} \rangle_{\rho} - \langle \hat A \rangle_{\rho} \langle \hat B \rangle_{\rho} \, . $ 
Using the definitions (\ref{G90a}) and (\ref{G90b}) and the hermiticity of $\hat \rho$, the expectation values of the cosine and sine operators are obtained as
\begin{eqnarray}
\langle \hat C \rangle_{\rho} \, &=& \, \sum_{n=0}^{\infty} f_{n} \, \Re (\rho_{n+1,n}) + \sum_{n=0}^{\infty} g_{n} \, \rho_{n,n} \, , \label{G300a} \\
\langle \hat S \rangle_{\rho} \, &=& \, \sum_{n=0}^{\infty} f_{n} \, \Im (\rho_{n+1,n}) + \sum_{n=0}^{\infty} g_{n} \, \rho_{n,n} \, , \label{G300b}
\end{eqnarray}
where $\Re(\Im)$ denotes the real (imaginary) part of a complex number. Compared with the expressions for $g_{n}=0$, the mean values for $g_{n}\neq 0$ are shifted by a state-dependent amount.
The expectation values of the operators squared are given by
\begin{eqnarray}
\langle \hat C^2 \rangle_{\rho} \, &=& \, 
\frac{1}{4} \sum_{n=0}^{\infty} (f_{n}^2+f_{n-1}^2+4g_{n}^2) \, \rho_{n,n} + 
\frac{1}{2} \sum_{n=0}^{\infty} f_{n} f_{n+1} \, \Re(\rho_{n+2,n}) \nonumber \\ 
&& + \sum_{n=0}^{\infty} f_{n} (g_{n}+g_{n+1}) \, \Re(\rho_{n+1,n}) , 
\label{G310a} \\
\langle \hat S^2 \rangle_{\rho} \, &=& \, 
\frac{1}{4} \sum_{n=0}^{\infty} (f_{n}^2+f_{n-1}^2+4g_{n}^2) \, \rho_{n,n} - 
\frac{1}{2} \sum_{n=0}^{\infty} f_{n} f_{n+1} \, \Re(\rho_{n+2,n}) \nonumber \\
&& + \sum_{n=0}^{\infty} f_{n} (g_{n}+g_{n+1}) \, \Im(\rho_{n+1,n}) , 
\label{G310b}
\end{eqnarray}
and those of the commutator and anticommutator by
\begin{eqnarray}
\langle {[\hat C , \hat S ]}_{-} \rangle_{\rho} \, &=& \, 
\frac{\rmi}{2} \sum_{n=0}^{\infty} (f_{n}^2 - f_{n-1}^2) \, \rho_{n,n} \nonumber \\
&& - \rmi \sum_{n=0}^{\infty} f_{n} (g_{n}-g_{n+1}) \, 
[\Re(\rho_{n+1,n}) + \Im(\rho_{n+1,n})] , \label{G310c} \\
\langle {[\hat C , \hat S ]}_{+} \rangle_{\rho} \, &=& \, 
2 \sum_{n=0}^{\infty} g_{n}^2 \, \rho_{n,n} +
\sum_{n=0}^{\infty} f_{n} f_{n+1} \, \Im(\rho_{n+2,n}) \nonumber \\
&& + \sum_{n=0}^{\infty} f_{n} (g_{n}+g_{n+1}) \, 
[\Re(\rho_{n+1,n}) + \Im(\rho_{n+1,n})] . \label{G310d}
\end{eqnarray}
We conclude the list of expectation values with those involving the number operator:
\begin{eqnarray}
\langle {[\hat N, \hat C]}_{-} \rangle_{\rho} &=& -\rmi \sum_{n=0}^{\infty} f_{n} \, \Im(\rho_{n+1,n}) , \label{G311a} \\
\langle {[\hat N, \hat S]}_{-} \rangle_{\rho} &=& \, \rmi \sum_{n=0}^{\infty} f_{n} \, \Re(\rho_{n+1,n}) , \label{G311b} \\
\langle {[\hat N, \hat C]}_{+} \rangle_{\rho} &=& \sum_{n=0}^{\infty} 
\Bigl[(2n+1) f_{n} \, \Re(\rho_{n+1,n}) + 2ng_{n} \, \rho_{n,n} \Bigr] , \label{G311c} \\
\langle {[\hat N, \hat S]}_{+} \rangle_{\rho} &=& \sum_{n=0}^{\infty} 
\Bigl[(2n+1) f_{n} \, \Im(\rho_{n+1,n}) + 2ng_{n} \, \rho_{n,n} \Bigr] . \label{G311d}
\end{eqnarray}
The expectation values considered so far involve only the neighbouring density matrix elements $\rho_{n,n}$, $\rho_{n+1,n}$ and $\rho_{n+2,n}$. This is a consequence of the tridiagonal structure of $\hat C$ and $\hat S$ and of having considered at most quadratic expressions in these operators.

Next we consider some uncertainty relations. The cosine-sine relation is
\begin{equation}
\fl
\sigma_{CC}^{\rho} \, \sigma_{SS}^{\rho}  \geq  
\frac{1}{16} \left( \sum_{n=0}^{\infty} \Big[ (f_{n}^2 - f_{n-1}^2) \, \rho_{n,n} - 
2 f_{n} (g_{n}-g_{n+1}) \, [\Re(\rho_{n+1,n}) + \Im(\rho_{n+1,n})] \Big] \right)^2 .
\end{equation}
Contrary to the Susskind--Glogower case, the right-hand-side contains not only the vacuum element $\rho_{0,0}$, but all diagonal elements $\rho_{n,n}$ and, if $g_{n}\neq 0$, even the off-diagonal elements $\rho_{n+1,n}$.
The number-cosine and number-sine uncertainty relations are
\begin{equation}
\fl
\sigma_{NN}^{\rho} \, \sigma_{CC}^{\rho} \geq 
\frac{1}{4} \, \left(\sum_{n=0}^{\infty} f_{n} \, \Im (\rho_{n+1,n})\right)^2 ,	\qquad \quad
\sigma_{NN}^{\rho} \, \sigma_{SS}^{\rho} \geq 
\frac{1}{4} \, \left(\sum_{n=0}^{\infty} f_{n} \, \Re (\rho_{n+1,n})\right)^2 .
\end{equation}
Here the right-hand-sides do not depend on $g_{n}$ and on the diagonal elements $\rho_{n,n}$.

We now give examples of expectation values for a Fock state $\vert n \rangle$. The cosine and sine operators have equal expectation values
\begin{equation}
\langle n \vert \hat C \vert n \rangle \, = \, \langle n \vert \hat S \vert n \rangle \, = \, g_{n} \, , \label{G370}
\end{equation}
which vanish for the generalized operators and yield a state-dependent shift (bias) for the extended operators. This is the expected result for a normal and a displaced oscillator, respectively. The expectation values of the quadratic operators are given by:
\begin{eqnarray}
\sigma_{CC}^{\vert n \rangle} \, = \, \sigma_{SS}^{\vert n \rangle} \, &=& \, 
\frac{1}{4} \, (f_{n}^2 + f_{n-1}^2) , \qquad \qquad
\sigma_{CS}^{\vert n \rangle} \, = \, 0  , \label{G380a} \\
\langle n \vert {[\hat C, \hat S]}_{-} \vert n \rangle \, &=& \, 
\frac{\rmi}{2} \, (f_{n}^2 - f_{n-1}^2) , \! \qquad \qquad
\langle n \vert {[\hat C, \hat S]}_{+} \vert n \rangle \, = \, 2 g_{n}^2 , \\
\langle n \vert \hat C^2 + \hat S^2 \vert n \rangle \, &=& \, 
\frac{1}{2} \, (f_{n}^2 + f_{n-1}^2 + 4 g_{n}^2) , \\
\langle n \vert {[\hat N, \hat C]}_{-} \vert n \rangle \, &=& \, 
\langle n \vert {[\hat N, \hat S]}_{-} \vert n \rangle \, = \, 0  , \qquad
\sigma_{NC}^{\vert n \rangle} \, = \, \sigma_{NS}^{\vert n \rangle} \, = \, 0  .
\end{eqnarray}
The cosine and sine operators are uncorrelated and their variances are equal and independent of $g_{n}$. 
The commutators of $\hat C$ and $\hat S$ with $\hat N$ have vanishing expectation values and the corresponding correlations vanish as well. These reasonable properties hold for both generalized and extended operators. If $g_{n}=0$ we have in addition 
$\langle n \vert \hat C^2 + \hat S^2 \vert n \rangle = 
2\, \sigma_{CC}^{\vert n \rangle}$,
i.e. the trigonometric relation $\langle n \vert \hat C^2 + \hat S^2 \vert n \rangle =1$ is violated to the same extent to which the variances squared deviate from their classical value $\frac{1}{2}$ (see (\ref{G650}) below). In the case of the classical polynomials and for large values of $n$ ($f_{n} \rightarrow 1$, $g_{n} \rightarrow 0$), the shifts of the mean values approach zero, the variances squared and 
$\langle n \vert \hat C^2 + \hat S^2 \vert n \rangle$
their classical values, and $\langle n \vert{[\hat C, \hat S]}_{-}\vert n \rangle$ vanishes. This is the expected behaviour on the basis of the correspondence principle.

\begin{figure}\begin{center}
\epsfbox{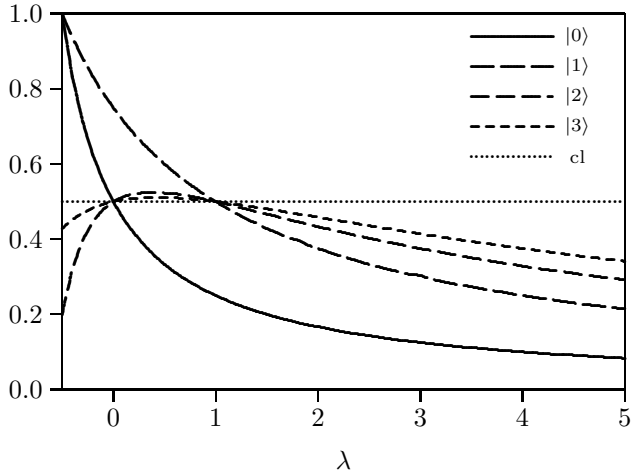}
\caption{\label{F1}Cosine and sine variances squared $\sigma_{CC}^{\vert n \rangle} = \sigma_{SS}^{\vert n \rangle}$ for the Fock states $\vert 0 \rangle$, $\vert 1 \rangle$, $\vert 2 \rangle$ and $\vert 3 \rangle$ as functions of $\lambda$, compared with the classical value $\frac{1}{2}$.}
\end{center}\end{figure}

We show in Figure~\ref{F1} the variances squared $\sigma_{CC}^{\vert n \rangle}=\sigma_{SS}^{\vert n \rangle}$ 
for the Fock states $\vert 0 \rangle$, $\vert 1 \rangle$, $\vert 2 \rangle$ and $\vert 3 \rangle$ as functions of the Gegenbauer parameter $\lambda$.
For $\lambda=1$ (Chebyshev polynomials of the second kind or Susskind--Glogower case) the variances squared are equal to the classical value $\frac{1}{2}$ for all Fock states $n\neq 0$ and to $\frac{1}{4}$ for $n = 0$. Similarly, for $\lambda \rightarrow 0$ (Chebyshev polynomials of the first kind) they are equal to $\frac{1}{2}$ for all $n\neq 1$ and to $\frac{3}{4}$ for $n = 1$. For $\lambda>1$ the variances squared stay below $\frac{1}{2}$ for all $n$, decrease to zero for very large $\lambda$ and increase with $n$ for fixed $\lambda$. For $n\ge2$ and $\lambda<0$ ($0<\lambda<1$) they stay below (above) $\frac{1}{2}$, with a maximum value at some point $\lambda_{\rm max}\in(0,1)$ which approaches $\frac{1}{2}$ for large $n$. Note that values of the variances squared greater (less) $\frac{1}{2}$ correspond to values of $\langle n \vert \hat C^2 + \hat S^2 \vert n \rangle$ greater (less) 1.

\begin{figure}\begin{center}
\epsfbox{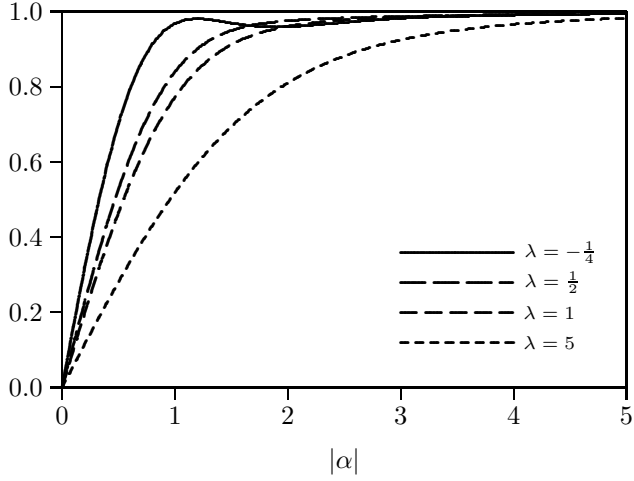}
\caption{\label{F2}$F_{1}(\vert \alpha \vert)$ for $\lambda = - \frac{1}{4}$, $\frac{1}{2}$, $1$ and $5$ as function of $\vert \alpha \vert$.}
\end{center}\end{figure}

As a next example we consider the coherent states 
\begin{equation}
\vert \alpha \rangle \, = \, \rme^{-\frac{\vert \alpha \vert^2}{2}} \, \sum_{n=0}^{\infty} \frac{\alpha^n}{\sqrt{n!}} \, \vert n \rangle , \qquad \quad 
(\alpha = \vert \alpha \vert \rme^{\rmi \varphi}).
\label{G400}
\end{equation}
The expectation values of the cosine and sine operators are now given by
\begin{equation}
\langle \alpha \vert \hat C \vert \alpha \rangle \, = \, F_{1} \, \cos \varphi + G_{1} \, , \qquad \ 
\langle \alpha \vert \hat S \vert \alpha \rangle \, = \, F_{1} \, \sin \varphi + G_{1} \, , \label{G410}
\end{equation}
and their variances squared and correlation by
\begin{eqnarray}
\fl \sigma_{CC}^{\vert\alpha\rangle} &=&  
(F_{2} - F_{1}^2) \cos^2 \varphi + \frac{1}{2} (F_{+} - F_{2}) + 
(G_{+} - 2F_{1}G_{1}) \cos\varphi \ + G_{2} - G_{1}^2 \, , \label{G411a} \\
\fl \sigma_{SS}^{\vert\alpha\rangle} &=& 
(F_{2} - F_{1}^2) \sin^2 \varphi \, + \frac{1}{2} (F_{+} - F_{2}) + 
(G_{+} - 2F_{1}G_{1}) \sin\varphi \ + G_{2} - G_{1}^2 \, , \label{G411b} \\
\fl \sigma_{CS}^{\vert \alpha \rangle} &=& 
(F_{2} - F_{1}^2) \cos \varphi \ \sin \varphi + \frac{1}{2}(G_{+} - 2F_{1}G_{1})(\cos\varphi + \sin\varphi) + G_{2} - G_{1}^2 \, . \label{G411c}
\end{eqnarray}
Here we have introduced the $\vert \alpha \vert$-dependent functions
\begin{eqnarray}
\{F_{1};\, G_{\pm}\} &=& \, \rme^{-\vert \alpha \vert^2} \sum_{n=0}^{\infty} 
\{f_{n}; \, f_{n}(g_{n}\pm g_{n+1})\} \, \frac{\vert \alpha \vert^{2n+1}}{n! \sqrt{n+1}} \, , \label{G412a} \\
F_{2} &=& \, \rme^{-\vert \alpha \vert^2} \sum_{n=0}^{\infty} f_{n}f_{n+1}\, \frac{\vert \alpha \vert^{2(n+1)}}{n! \sqrt{(n+1)(n+2)}} \, , \label{G412b} \\ 
\{F_{\pm}; \, G_{1}; \, G_{2}\} &=& \, \rme^{-\vert \alpha \vert^2} \sum_{n=0}^{\infty} \{\frac{1}{2} (f_{n}^2 \pm f_{n-1}^2 ); \, g_{n}; \, g_{n}^2 \} \, \frac{\vert \alpha \vert^{2n}}{n!} \, . \label{G412c} 
\end{eqnarray}
Some other expectation values of interest are:
\begin{eqnarray}
\langle \alpha \vert {[\hat C, \hat S]}_{-} \vert \alpha \rangle \, &=& \, 
\rmi \, [F_{-} - G_{-} (\cos \varphi + \sin \varphi)] , \label{G415a} \\ 
\langle \alpha \vert \hat C^2 + \hat S^2 \vert \alpha \rangle \, &=& \, 
F_{+} + 2\,G_{2} + G_{+} (\cos \varphi + \sin \varphi) , \label{G415b} \\ 
\langle \alpha \vert {[\hat N, \hat C]}_{-} \vert \alpha \rangle \, &=&  
-\rmi \, F_{1} \sin \varphi , \qquad \ 
\langle \alpha \vert {[\hat N, \hat S]}_{-} \vert \alpha \rangle \, = \, 
\rmi \, F_{1} \cos \varphi . \label{G415c}
\end{eqnarray}
For the generalized operators all $G$ functions vanish and the relations above take on familiar forms. These forms get usually modified for the extended operators  by state-dependent terms. Thus, the expectation values of $\hat C$ and $\hat S$ get shifted by the same amount $G_{1} (\vert\alpha\vert)$. 
For the classical polynomials and for large values of $\vert \alpha \vert$ we recover the classical results
$\langle\alpha\vert \hat C \vert\alpha\rangle \rightarrow \cos \varphi$ and
$\langle\alpha\vert \hat S \vert\alpha\rangle \rightarrow \sin \varphi$, 
since $F_{1} \rightarrow 1$ and $G_{1} \rightarrow 0$ in this limit. Figure~\ref{F2} displays $F_{1}(\vert \alpha \vert)$ as a function of $\vert\alpha\vert$ for 
$\lambda = - \frac{1}{4}$, $\frac{1}{2}$, $1$ and $5$. 
The functions increase monotonically with $\vert \alpha \vert$ for $\lambda > 0$ and 
develop a bump followed by a valley for $\lambda < 0$. All functions start at zero for $\vert\alpha\vert =0$ (vacuum state) and approach $1$ (classical limit) for $\vert\alpha\vert\to\infty$; the approach is slower for higher values of $\lambda$. 

\section{Cosine and sine representations and distributions} \label{S4}

In this section we introduce the cosine and sine representations of an arbitrary state and discuss the corresponding probability distributions. The cosine and sine distributions are considered on their own right as functions of the eigenvalues $c$ and $s$, without expressing the eigenvalues in terms of phases.\footnote[1]{In the definitions (\ref{G90a}) ((\ref{G90b})) of the operators $\hat C$ ($\hat S$) there is no reference to a phase and labeling the eigenvalues $c$ ($s$) by $\cos \theta$ ($\sin \theta$) should be considered merely as a parameterization at this stage.} The importance of this kind of distributions for the quantum measurement of trigonometric variables has been stressed in \cite{ar}.

The cosine and sine states (\ref{G96}) can be used to define cosine and sine representations of an arbitrary normalized state $\vert \psi \rangle$ with Fock components $\psi_{n}=\langle n \vert \psi \rangle$ by
\begin{equation}
\fl
\psi^{(\rm C)}(c) \equiv \langle c \vert \psi \rangle = \sum_{n=0}^{\infty} p_{n}(c) \, \psi_{n} , \qquad \qquad
\psi^{(\rm S)}(s) \equiv \langle s \vert \psi \rangle = \sum_{n=0}^{\infty} (-\rmi)^n \, p_{n}(s) \, \psi_{n} . 
\label{G600}
\end{equation}
The cosine and sine distributions are then given by 
$\vert \psi^{(\rm C)}(c) \vert^2$ and $\vert \psi^{(\rm S)}(s) \vert^2$ or, more generally, for a mixed state with density matrix elements $\rho _{m,n}$ by
\begin{equation}
\fl
\mathcal{P}_{\rho}^{(\rm C)}(c) = 
\sum_{m,n=0}^{\infty} \rho_{mn} \, p_{m}(c) p_{n}(c) , \qquad \qquad
\mathcal{P}_{\rho}^{(\rm S)}(s) = 
\sum_{m,n=0}^{\infty} (- \rmi)^{m-n} \, \rho_{mn} \, p_{m}(s) p_{n}(s) .
\label{G620}
\end{equation}
Since $p_{m}(x) p_{n}(x)$ is symmetric in $m$ and $n$, only the symmetric part of the remaining factors contribute to the double sums. Taking the hermiticity of the density matrix into account, this symmetric part is 
$\Re [\rho_{m,n}]$ for $x=c$ and $\Re [(- \rmi)^{m-n} \rho_{m,n}]$ for $x=s$. 
The cosine and sine probability distributions are then given by
\begin{eqnarray}	 
\mathcal{P}_{\rho}^{(\rm C)}(c) &=& \sum_{n=0}^{\infty}   
\rho_{n,n} \, p_{n}^2(c) +  
2 \sum_{m>n=0}^{\infty} \Re[\rho_{m,n}]\, p_{m}(c) p_{n}(c) , \label{G624} \\
\mathcal{P}_{\rho}^{(\rm S)}(s) &=& \sum_{n=0}^{\infty}  
\rho_{n,n} \, p_{n}^2(s) +  
2 \sum_{m>n=0}^{\infty} \Re[\rho_{m,n} \, \rme^{-\rmi \frac{\pi}{2}(m-n)}] \, p_{m}(s) p_{n}(s) . 
\label{G625}
\end{eqnarray}
The sine distributions follow formally from the cosine distributions by replacing the real part of the density matrix elements $\rho_{m,n}$ by the real part of the phase-shifted matrix elements $\rho_{m,n} \exp[-\rmi \frac{\pi}{2}(m-n)]$. 
For a large class of states, including the coherent and squeezed states, the density matrix elements $\rho_{m,n}$ have a phase dependence of the form $\exp[\rmi (m-n) \varphi]$. In this case the sine distributions follow from the cosine distributions by the substitutions $(c,\varphi) \rightarrow (s,\varphi - \frac{\pi}{2})$.
In other words, the phases `measured' by the cosine and sine distributions are in quadrature as expected.

The cosine and sine distributions can be used to calculate the expectation values of functions $F(\hat C)$ and $F(\hat S)$ 
according to
\begin{equation}
\langle F(\hat C) \rangle_{\rho} = \int_{-1}^{+1} \rmd c \, F(c) \, \mathcal{P}_{\rho}^{(\rm C)}(c) , \qquad
\langle F(\hat S) \rangle_{\rho} = \int_{-1}^{+1} \rmd s \, F(s) \, \mathcal{P}_{\rho}^{(\rm S)}(s) .
\label{G630} 
\end{equation}

\begin{figure}\begin{center}
\epsfbox{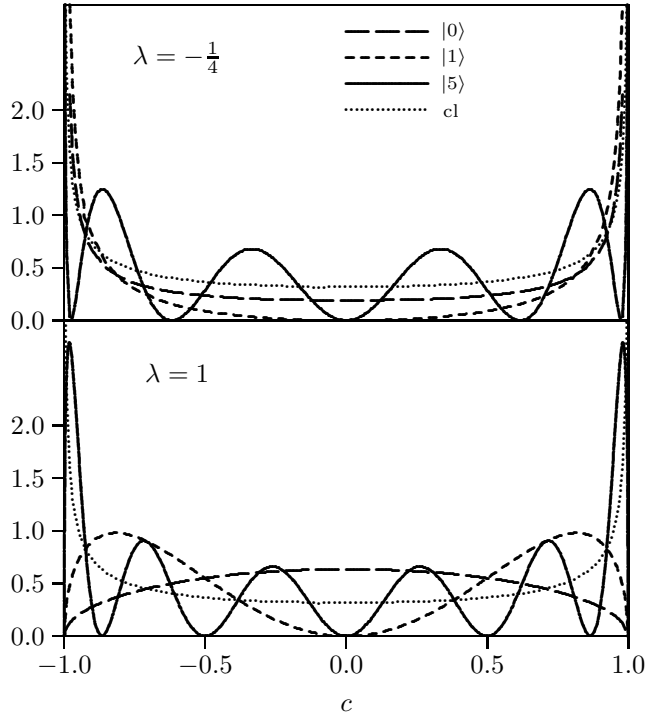}
\caption{\label{F3}Cosine (and equal sine) distributions $\mathcal{P}_{\vert n \rangle}^{(\rm C)}(c)$ for the Fock states $\vert 0 \rangle$, $\vert 1 \rangle$ and $\vert 5 \rangle$, for $\lambda = - \frac{1}{4}$ and $1$, compared with the classical distribution $\mathcal{P}_{\rm cl}^{(\rm C)}(c)$.}
\end{center}\end{figure}

We now apply the above considerations to the Fock and coherent states. Using (\ref{G96}), the cosine and sine representations of a Fock state $\vert n \rangle$ are given by 
\begin{equation}
\langle c \vert n \rangle = p_{n}(c) , \qquad \qquad \qquad
\langle s \vert n \rangle = (-\rmi)^n \, p_{n}(s) , \label{G635}
\end{equation}
i.e. by the same function $p_{n}(x)$ evaluated for $x=c$ and $x=s$, respectively, with an additional factor $(-\rmi)^n$ for $\langle s \vert n \rangle$.\footnote[1]{This is in perfect analogy with the usual coordinate ($\langle q \vert n \rangle = h_{n}(q)$) and momentum ($\langle p \vert n \rangle = (-\rmi)^n h_{n}(p)$) representations of the $n$th energy eigenstate of the harmonic oscillator.}
The corresponding cosine and sine distributions are equal and given by
\begin{equation}
\mathcal{P}_{\vert n \rangle}^{(\rm X)}(x) = p_{n}^2(x) = 
w(x) \, P_{n}^2(x) / {d_{n}} \, , \qquad (x=c, s).
\label{G640} 
\end{equation}
This is precisely the expression under the normalization integral (\ref{G56}) for $n=m$, now interpreted as the probability distribution to find the value $x$ in the Fock state $\vert n \rangle$. The distribution for the vacuum state is given by the normalized weight function $w(x)/\mu_{0}$, where $\mu_{0}$ is the zeroth moment (\ref{G37}); if $P_{0}(x)=1$ then $\mu_{0}=d_{0}$. 
The distribution (\ref{G640}) vanishes at the $n$ zeros of the polynomial $P_{n}(x)$, $n \geq 1$, all of which are known \cite{sz} to be real, distinct and located in the interior of the interval $\mbox{[-1,1]}$. The distribution therefore shows an oscillatory behaviour with $n+1$ peaks.
It is symmetric for the generalized and asymmetric for the extended operators. 
The behaviour near the end points $x=\pm 1$ depends solely on the value of $w(\pm 1)$, since $P_{n}(\pm 1)$ is finite.
In the particular case of the Jacobi polynomials the distributions are asymmetric (symmetric) for $\mu \neq \nu$ ($\mu = \nu$). 
The behaviour near the end points is given by $(1-x)^{\mu}$ for $x\rightarrow 1$ and by $(1+x)^{\nu}$ for $x\rightarrow -1$.
Due to the property 
$p_{n}^{(\mu,\nu)}(-x) = (-1)^n \, p_{n}^{(\nu,\mu)}(x)$,
the distributions with Jacobi indices $(\mu,\nu)$ and $(\nu,\mu)$ are related by reflection. The vacuum state distributions are given by $w(x)/d_{0}$, with $d_{0}$ from the Tables~\ref{T1} and ~\ref{T2}. They are uniform for the Legendre polynomial $P_{0}(x)$ and equal to the classical distribution 
\begin{equation}
\mathcal{P}_{\rm cl}^{(\rm X)}(x) = \frac{1}{\pi \sqrt{1-x^2}} \qquad \qquad(x=c, s) 
\label{G650}
\end{equation}
for the Chebyshev polynomial $T_{0}(x)$. The distribution (\ref{G650}) corresponds to a uniform arccosine or arcsine distribution (see section \ref{S5}), yields vanishing expectation values and variances squared equal to $\frac{1}{2}$.

\begin{figure}\begin{center}
\epsfbox{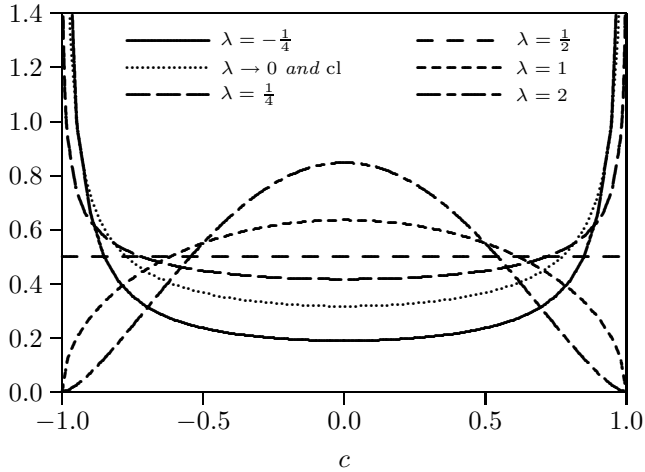}
\caption{\label{F5}Cosine (and equal sine) distributions 
$\mathcal{P}_{\vert 0 \rangle}^{(\rm C)}(c)$ for the vacuum state $\vert 0 \rangle$, for $\lambda=-\frac{1}{4}$, $\rightarrow 0$, $\frac{1}{4}$, $\frac{1}{2}$, $1$ and $2$, compared with the classical distribution $\mathcal{P}_{\rm cl}^{(\rm C)}(c)$.}
\end{center}\end{figure}

\begin{figure}\begin{center}
\epsfbox{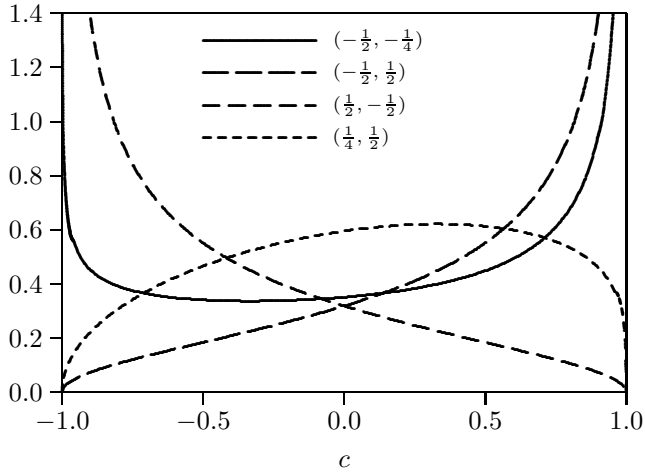}
\caption{\label{F14}Cosine (and equal sine) distributions 
$\mathcal{P}_{\vert 0 \rangle}^{(\rm C)}(c)$ for the vacuum state $\vert 0 \rangle$ and for the Jacobi parameter values $(\mu,\nu)=(-\frac{1}{2},-\frac{1}{4}), (-\frac{1}{2},\frac{1}{2}), (\frac{1}{2},-\frac{1}{2}), (\frac{1}{4},\frac{1}{2})$.}
\end{center}\end{figure}

The general behaviour outlined above can be clearly observed in the figures~\ref{F3} to \ref{F14}, where the classical distribution is displayed for comparison. 
Figure~\ref{F3} shows the cosine (and equal sine) distributions for the Fock states $\vert 0 \rangle$, $\vert 1 \rangle$ and $\vert 5 \rangle$ and the Gegenbauer parameters $\lambda = -\frac{1}{4}$, $1$. The distributions for $\lambda = -\frac{1}{4}$ ($\lambda = 1$) diverge (vanish) at the end points and have $n$ ($n+2$) zeros and $n+1$ peaks for $n\geq1$. The quantum distributions oscillate around the classical distribution, approaching it for large values of $n$.\footnote {This is reminiscent of the coordinate and momentum distributions for the $n$th 
eigenstate of the harmonic oscillator, in which case the zeros of the distributions are the roots of the Hermite polynomials.}


Figure~\ref{F5} shows the cosine (and equal sine) distributions for the vacuum state $\vert 0 \rangle$ and for different values of $\lambda$: $ -\frac{1}{4}$, $0$,  $\frac{1}{4}$, $\frac{1}{2}$, $1$  and $2$. For $\lambda > \frac{1}{2}$ ($\lambda < \frac{1}{2}$) the distributions vanish (diverge) at the end points $c=\pm 1$ and have a maximum (minimum) increasing with $\lambda$ at $c=0$.

Figure~\ref{F14} shows the cosine (and equal sine) distributions for the vacuum state $\vert 0 \rangle$ and various Jacobi parameter sets $(\mu,\nu)$. The distributions are asymmetric and show the behaviour near the end points as discussed above. The asymmetry of the vacuum state distributions is at first glance an unexpected result, but it may be natural for a `displaced' oscillator ($\mu \neq \nu$). By contrast, the distributions in Figure~\ref{F5} corresponding to a  `normal' oscillator ($\mu = \nu$) show the expected symmetry.

We now come to the coherent states (\ref{G400}). 
Their cosine and sine representations are
\begin{equation}
\fl
\langle c \vert \alpha \rangle = \, \rme^{-\frac{\vert \alpha \vert^2}{2}} \, \sum_{n=0}^{\infty} \frac{\vert \alpha \vert ^n}{\sqrt{n!}} \, 
\rme^{\rmi n \varphi} \, p_{n}(c) , \qquad \quad
\langle s \vert \alpha \rangle = \, \rme^{-\frac{\vert \alpha \vert^2}{2}} \, \sum_{n=0}^{\infty} \frac{\vert \alpha \vert ^n}{\sqrt{n!}} \, 
\rme^{\rmi n (\varphi - \frac{\pi}{2})} \, p_{n}(s) .
\label{G800}
\end{equation}
The cosine probability distribution is then given by
\begin{equation}
\fl \mathcal{P}_{\vert \alpha \rangle}^{(\rm C)}(c) \, = \, 
\rme^{- \vert \alpha \vert^2} \, \sum_{n=0}^{\infty} \, 
\Bigl\{\frac{\vert \alpha \vert^{2n}}{n!}\, p_{n}^2(c) +  
2 \sum_{m(>n)}^{\infty} \frac{\vert \alpha \vert^{m+n}}
{\sqrt{m!\, n!}} \, \cos[(m-n) \varphi] \, p_{m}(c) p_{n}(c) \Bigr\} , 
\label{G840}
\end{equation}
and the sine 
distribution $\mathcal{P}_{\vert \alpha \rangle}^{(\rm S)}(s)$ follows therefrom by replacing $(c, \varphi) \rightarrow (s, \varphi - \frac{\pi}{2})$ on the right-hand-side.
The dependence on the coherent phase is through $\cos[(m-n) \varphi]$ ($\cos[(m-n) (\varphi- \frac{\pi}{2})]$) for the cosine (sine) distribution. For a fixed phase $\varphi_{0}$, the states with $\varphi = \varphi_{0}$ and $\varphi = - \varphi_{0}$ ($\varphi = \varphi_{0}$ and $\varphi = \pi - \varphi_{0}$) then have the same cosine (sine) distribution. These are precisely the phases which yield the same value of the cosine ($c=\cos \varphi_{0}$) or sine ($s=\sin \varphi_{0}$). States having identical cosine (sine) distributions are distinguished by their sine (cosine) distributions, so that both distributions are needed. Note also that the sine distributions for the states with $\varphi = \frac{\pi}{2} \pm \varphi_{0}$ are identical with the cosine distribution for the state with $\varphi = \varphi_{0}$.

\begin{figure}\begin{center}
\epsfbox{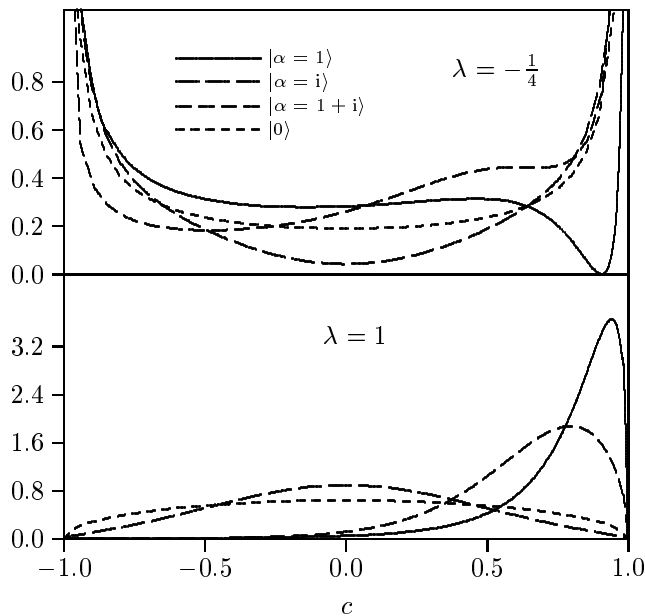}
\caption{\label{F6}Cosine distributions 
$\mathcal{P}_{\vert \alpha \rangle}^{(\rm C)}(c)$ 
for the coherent states with 
$\alpha = 1$, $\rmi$, $1+\rmi$, $0$, 
and for $\lambda=-\frac{1}{4}$, $1$. Identical distributions for 
$\alpha = 1$, $-\rmi$, $1-\rmi$, $0$, respectively.}
\end{center}\end{figure}

\begin{figure}\begin{center}
\epsfbox{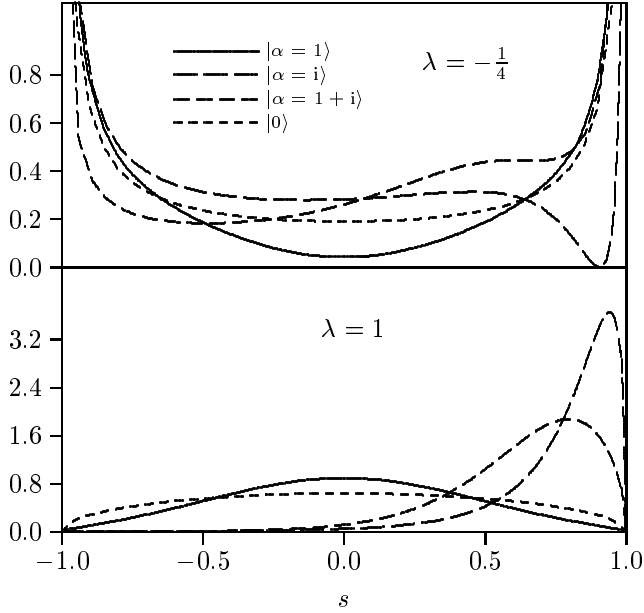}
\caption{\label{F16}Sine distributions 
$\mathcal{P}_{\vert \alpha \rangle}^{(\rm S)}(s)$ 
for the coherent states with 
$\alpha = 1$, $\rmi$, $1+\rmi$, $0$,
and for $\lambda=-\frac{1}{4}$, $1$. Identical distributions for 
$\alpha = -1$, $\rmi$, $-1+\rmi$, $0$, respectively.}
\end{center}\end{figure}

Figure~\ref{F6} shows the cosine distributions for $\lambda=-\frac{1}{4}$, $1$ of the coherent states with $\alpha=1$, $\rmi$, $1+\rmi$, $0$ (vacuum state). The curves referring to $\alpha=\rmi$, $1+\rmi$ give also the cosine distributions for the complex conjugate values $\alpha=-\rmi$, $1-\rmi$.
The sine distributions for the same input data are shown in Figure~\ref{F16}. Here the curves for $\alpha=1$, $1+\rmi$ refer also to the sine distributions for $\alpha=-1$, $-1+\rmi$, respectively. 
The cosine distributions for $\alpha=1$, $\rmi$, $1+\rmi$, $0$ and the sine distributions for $\alpha=\rmi$, $\pm 1$, $\pm 1+\rmi$, $0$ are identical in that order.
The distributions for $\lambda \geq 0$ peak, as expected, around the cosine and sine values corresponding to the coherent phase, $\varphi = 0$ ($c=1$, $s=0$), $\varphi = \frac{\pi}{2}$ ($c=0$, $s=1$) and $\varphi = \frac{\pi}{4}$ ($c=s \approx 0.71$), 
but develop unnatural dips for $\lambda<0$. The latter distributions should be considered therefore with care. The peaks of the distributions 
get sharper with increasing values of $\vert \alpha \vert$.

\section{Arccosine and arcsine operators, representations and distributions} \label{S5}

In this section we consider the inverse trigonometric operators and use their eigenstates to define arccosine and arcsine representations and corresponding probability distributions. The direct sampling of such distributions in the particular Susskind--Glogower case has been considered in \cite{dkw}.

The arccosine and arcsine operators 
are defined by the power-series expansions 
\begin{eqnarray}
\it \hat \Theta_{c} &\equiv& \arccos \hat C \, = \, \frac{\pi}{2} \, \hat 1 - \sum_{k=0}^{\infty} \, \frac{(-1)^k}{2k+1} \, 
\Big( \begin{array}{c}\! - \frac{1}{2} \\ k \end{array} \! \Big) \, \hat C^{2k+1} , \label{G900a} \\
\it \hat \Theta_{s} &\equiv& \arcsin \hat S \ = \, \sum_{k=0}^{\infty} \, \frac{(-1)^k}{2k+1} \, 
\Big( \begin{array}{c}\! - \frac{1}{2} \\ k \end{array} \! \Big) \, \hat S^{2k+1} . \label{G900b}
\end{eqnarray}
They are of the generalized or extended type depending on the type of the corresponding trigonometric operator. The series are well defined even for the extended operators, since the spectrum of the latter is also confined to the convergence intervall $\mbox{[-1,1]}$. Combining (\ref{G900a}) and (\ref{G900b}) we obtain $\arccos \hat C = \frac{\pi}{2} \hat 1 - \arcsin \hat C$ and $\arccos \hat S = \frac{\pi}{2} \hat 1 - \arcsin \hat S$, which are the quantum versions of the classical relation $\arccos x = \frac{\pi}{2} - \arcsin x$.

The operators $\it \hat \Theta_{c}$ and $\it \hat \Theta_{s}$ do not commute, but 
$\it \hat \Theta_{c}$ commutes with $\hat C$ and $\it \hat \Theta_{s}$ with $\hat S$.
It then follows that the states $\vert c \rangle$ and $\vert s \rangle$ are also eigenstates of $\it \hat \Theta_{c}$ and $\it \hat \Theta_{s}$ 
\begin{equation}
\it \hat \Theta_{c} \, \vert c \rangle = \arccos c \ \vert c \rangle , \qquad \qquad \quad \ \ \it \hat \Theta_{s} \, \vert s \rangle = \arcsin s \ \vert s \rangle .
\end{equation}
Writing $c=\cos \theta_{c}$ and $s=\sin \theta_{s}$, and choosing the principal value branches $0 \le \theta_{c} \le \pi$ and $-\frac{\pi}{2} \le \theta_{s} \le +\frac{\pi}{2}$, we define the arccosine and arcsine states by
\begin{eqnarray}
\vert \theta_{c} \rangle &\equiv& \sqrt{\sin \theta_{c}} \, [\, \vert c \rangle \, ]_{c = \cos \theta_{c}} \, = \, \sum_{n=0}^{\infty}  c_{n}(\theta_{c}) \, \vert n \rangle , 
\label{G910a} \\
\vert \theta_{s} \rangle &\equiv& \sqrt{\cos \theta_{s}} \, [\, \vert s \rangle \, ]_{s = \sin \theta_{s}} \, = \, \sum_{n=0}^{\infty} \rmi^n \, s_{n}(\theta_{s}) \, \vert n \rangle , 
\label{G910b}
\end{eqnarray}
with $c_{n}(\theta_{c})$ and $s_{n}(\theta_{s})$ given in (\ref{G71}) and (\ref{G72}).  
The states $\vert \theta_{c} \rangle$ ($\vert \theta_{s} \rangle$) 
are eigenstates of $\it \hat \Theta_{c}$ ($\it \hat \Theta_{s}$) 
as well as of $\hat C$ ($\hat S$) with appropriate eigenvalues: 
\begin{eqnarray}
\it \hat \Theta_{c} \, \vert \theta_{c} \rangle &=& \ \theta_{c} \, \vert \theta_{c} \rangle , \label{G911} \qquad \quad \ \qquad \qquad
\it \hat \Theta_{s} \, \vert \theta_{s} \rangle = \ \theta_{s} \, \vert \theta_{s} \rangle , \label{G912} \\
\hat C \, \vert \theta_{c} \rangle &=& \, \cos \theta_{c} \, \vert \theta_{c} \rangle , \qquad \! \qquad \qquad
\hat S \, \vert \theta_{s} \rangle \ \, = \ \sin \theta_{s} \, \vert \theta_{s} \rangle . 
\end{eqnarray}
They are orthogonal and yield a resolution of unity according to 
\begin{equation}
\langle \theta_{c} \vert \theta_{c}' \rangle = \delta (\theta_{c} - \theta_{c}') , \qquad \qquad \qquad \,  
\langle \theta_{s} \vert \theta_{s}' \rangle = \delta (\theta_{s} - \theta_{s}') , \label{G920}
\end{equation}
\begin{equation}
\int_{0}^{\pi} \rmd \theta_{c} \, \vert \theta_{c} \rangle \langle \theta_{c} \vert \, = \, \hat 1 , \qquad \qquad \qquad 
\int_{-\pi /2}^{+\pi /2} \rmd  \theta_{s} \, \vert \theta_{s} \rangle \langle \theta_{s} \vert \, = \, \hat 1 . \label{G930} 
\end{equation}
The inverse trigonometric operators and their eigenstates are related by
\begin{equation}
{\it \hat \Theta_{c}} = \rme^{- \rmi \frac{\pi}{2} \hat N} 
\left[ \frac{\pi}{2}  \hat 1 - \it \hat \Theta_{s} \right] \rme^{+ \rmi \frac{\pi}{2} \hat N} , \qquad 
{\it \hat \Theta_{s}} = \rme^{+ \rmi \frac{\pi}{2} \hat N}  
\left[ \frac{\pi}{2}  \hat 1 - \it \hat \Theta_{c} \right] \rme^{- \rmi \frac{\pi}{2} \hat N} , 
\label{G932}
\end{equation}
\begin{equation}
\vert \theta_{c} \rangle \, = \, [\, \rme^{- \rmi \frac{\pi}{2} \hat N} \,  \vert \theta_{s} \rangle \, ]_{\theta_{s}=\frac{\pi}{2} - \theta_{c}} \, ,\, \qquad \quad
\vert \theta_{s} \rangle \, = \, [\, \rme^{+ \rmi \frac{\pi}{2} \hat N} \,  \vert \theta_{c} \rangle \, ]_{\theta_{c}= \frac{\pi}{2} - \theta_{s}}  \, . 
\label{G933}	
\end{equation}
Hence, $\it \hat \Theta_{s}$ ($\it \hat \Theta_{c}$) is unitarily related to $[\frac{\pi}{2} \, \hat 1 - \it \hat \Theta_{c} ]$ 
([$\frac{\pi}{2} \, \hat 1 - \it \hat \Theta_{s} ]$) 
and not to $\it \hat \Theta_{c}$ ($\it \hat \Theta_{s}$) itself.

We have called the states (\ref{G910a}) and (\ref{G910b}) the arccosine and arcsine states, respectively. The terms cosine-phase and sine-phase states would also be appropriate and are actually used for the well-known Susskind--Glogower states
\begin{equation}
\fl
\vert \theta_{c} \rangle_{\rm SG} = \sqrt{\frac{2}{\pi}} \sum_{n=0}^{\infty} \sin [(n+1)\theta_{c}] \, \vert n \rangle , \qquad
\vert \theta_{s} \rangle_{\rm SG} = \sqrt{\frac{2}{\pi}} \sum_{n=0}^{\infty} \rmi^n \sin[(n+1)(\frac{\pi}{2}-\theta_{s})] \, \vert n \rangle ,
\label{G935} 
\end{equation}
to which our states reduce in the particular case of the Chebyshev polynomials of the second kind by using the representation $U_{n}(\cos \theta) = \sin [(n+1)\, \theta \, ] / \sin \theta$. 
The states (\ref{G935}) are usually denoted by $\vert \cos \theta_{c} \rangle$ ($\vert \sin \theta_{s} \rangle$) to emphasize the eigenvalues of $\hat C$ ($\hat S$) rather than of $\it \hat \Theta_{c}$ ($\it \hat \Theta_{s}$). Nevertheless, they are normalized to $\delta (\theta - \theta')$ and the measure in the resolution of unity is $\rmd \theta$, as in our case. From this point of view our notation is more consistent with Dirac's convention. Also, in the present context the notations $\vert \cos \theta_{c} \rangle$ ($\, \vert \sin \theta_{s} \rangle$) for the states (\ref{G910a}) ((\ref{G910b})) could give rise to confusion with the only reparameterized states $[\vert c \rangle]_{c=\cos \theta_{c}}$ ($[\vert s \rangle]_{s=\sin \theta_{s}}$) not including the square root of the Jacobian.
Note that $\theta_{c}$ ($\theta_{s}$) are not true phases but, strictly speaking, the arccosine (arcsine) of $c$ ($s$) and vary accordingly over a $\pi$-range instead of a $2\pi$-range characteristic for a genuine phase.

The states $\vert \theta_{c} \rangle$ ($\vert \theta_{s} \rangle$) in (\ref{G910a})((\ref{G910b})) have the same structure as the states $\vert c \rangle$ ($\vert s \rangle$) in (\ref{G96}), with $c_{n}(\theta_{c})$ ($s_{n}(\theta_{s})$) replacing $p_{n}(c)$ ($p_{n}(s)$). These replacements act as a rule when switching between appropriate trigonometric and inverse trigonometric quantities. Thus, 
the arccosine and arcsine (or: cosine-phase and sine-phase) representations of a state $\vert \psi \rangle$ are given by (compare (\ref{G600}))
\begin{equation}
\fl
\psi^{(\Theta_{c})}(\theta_{c}) \equiv \langle \theta_{c} \vert \psi \rangle = \sum_{n=0}^{\infty} c_{n}(\theta_{c}) \, \psi_{n} ,	\qquad  \quad
\psi^{(\Theta_{s})}(\theta_{s}) \equiv \langle \theta_{s} \vert \psi \rangle = 
\sum_{n=0}^{\infty} (-\rmi)^n \, s_{n}(\theta_{s}) \, \psi_{n} .
\label{G940}
\end{equation}
The corresponding arccosine and arcsine (or: cosine-phase and sine-phase) probability distributions follow from (\ref{G624}) and (\ref{G625}) by the same replacements
\begin{equation}
\mathcal{P}_{\rho}^{(\Theta_{c})}(\theta_{c}) = [\mathcal{P}_{\rho}^{(\rm C)}(c)]_{p_{n}(c) \rightarrow c_{n}(\theta_{c})}, \qquad 	
\mathcal{P}_{\rho}^{(\Theta_{s})}(\theta_{s}) = [\mathcal{P}_{\rho}^{(\rm S)}(s)]_{p_{n}(s) \rightarrow s_{n}(\theta_{s})}.
\label{G950}
\end{equation}
For the expectation values of functions $F(\it \hat \Theta_{c})$ and $F(\it \hat \Theta_{s})$ we then have
\begin{eqnarray}
\langle F(\it \hat \Theta_{c}) \rangle_{\rho} &=& 
\int_{0}^{\pi} \rmd \theta_{c} \, F(\theta_{c}) \, \mathcal{P}_{\rho}^{(\Theta_{c})}(\theta_{c}) \quad \ = \int_{-1}^{+1} \rmd c \, 
F(\arccos c) \, \mathcal{P}_{\rho}^{(\rm C)}(c) , 
\label{G960a} \\
\langle F(\it \hat \Theta_{s}) \rangle_{\rho} &=& \int_{-\pi /2}^{+\pi /2} \rmd \theta_{s} \, F(\theta_{s}) \, \mathcal{P}_{\rho}^{(\Theta_{s})}(\theta_{s}) = \int_{-1}^{+1} \rmd s \, F(\arcsin s) \, \mathcal{P}_{\rho}^{(\rm S)}(s) . 
\label{G960b}
\end{eqnarray}
These relations 
reduce to (\ref{G630}) if $F$ depends on $\it \hat \Theta_{c}$ ($\it \hat \Theta_{s}$) via $\cos \it \hat \Theta_{c}=\hat C$ ($\sin \it \hat \Theta_{s}=\hat S$) only.
The classical uniform distributions correspond to
\begin{equation}
\mathcal{P}_{\rm cl}^{(\Theta_{c})}(\theta_{c})=\frac{1}{\pi} \, \quad \mathrm{on} \ [0,\pi], \qquad \quad
\mathcal{P}_{\rm cl}^{(\Theta_{s})}(\theta_{s})=\frac{1}{\pi} \quad \mathrm{on} \ [-\frac{\pi}{2},\frac{\pi}{2}] . \label{G961}	
\end{equation}
They yield the expectation values 
$\langle \theta_{c} \rangle_{\rm cl} = \frac{\pi}{2}$ and 
$\langle \theta_{s} \rangle_{\rm cl} = 0$, 
equal variances squared   
$\sigma_{\theta_{c}\theta_{c}}^{\rm cl} = 
\sigma_{\theta_{s}\theta_{s}}^{\rm cl} = \pi^2/12$
and the classical distribution (\ref{G650}) for $x=\cos\theta_{c}$ and $x=\sin\theta_{s}$.

The arccosine and arcsine representations of a Fock state $\vert n \rangle$ are given by
\begin{equation}
\langle \theta_{c} \vert n \rangle = \, c_{n}(\theta_{c}), \qquad \qquad \qquad \ \ 
\langle \theta_{s} \vert n \rangle = \, (-\rmi)^n \, s_{n}(\theta_{s}),
\end{equation}
and yield the probability distributions (related by $\theta_{c}=\frac{\pi}{2}-\theta_{s}$) 
\begin{equation}
\mathcal{P}_{\vert n \rangle}^{(\Theta_{c})}(\theta_{c}) \, = \, c_{n}^2(\theta_{c}),  \qquad \qquad \ \quad 
\mathcal{P}_{\vert n \rangle}^{(\Theta_{s})}(\theta_{s}) \, = \, s_{n}^2(\theta_{s}).  
\label{G962} 	
\end{equation}
Using (\ref{G932}) and the binomial expansion formula, 
we obtain 
the following relation between the arccosine and arcsine moments
\begin{equation}
\langle n \vert {\it \hat \Theta_{c}}^{k} \vert n \rangle = \Bigl(\frac{\pi}{2}\Bigr)^{k} + 
\sum_{l=1}^{k} (-1)^{l} \Bigl( \! \begin{array}{c} k \\ l \end{array} \! \Bigr)
\Bigl(\frac{\pi}{2}\Bigr)^{k-l} \, 
\langle n \vert {\it \hat \Theta_{s}}^{l} \vert n \rangle .
\label{G964}
\end{equation}
In particular, for $k=1$ and $k=2$ we find 
\begin{equation}
\fl
\langle n \vert {\it \hat \Theta_{c}} \vert n \rangle = \frac{\pi}{2} - 
\langle n \vert {\it \hat \Theta_{s}} \vert n \rangle ,	\qquad \qquad
\langle n \vert {{\it \hat \Theta_{c}}}^{2} \vert n \rangle = \Bigl(\frac{\pi}{2}\Bigr)^{2} - \pi \, \langle n \vert {\it \hat \Theta_{s}} \vert n \rangle + \langle n \vert {{\it \hat \Theta_{s}}}^{2} \vert n \rangle ,
\label{G966}
\end{equation}
implying equal variances squared, 
$\sigma_{\it \Theta_{c} \it \Theta_{c}}^{\vert n \rangle} =  \sigma_{\it \Theta_{s} \it \Theta_{s}}^{\vert n \rangle}$.
\begin{figure}\begin{center}
\epsfbox{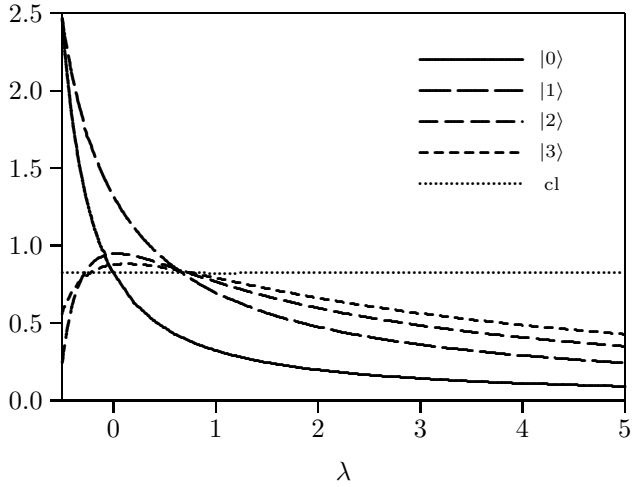}
\caption{\label{F7}Arccosine and arcsine variances squared $\sigma_{\it \Theta_{c} \it \Theta_{c}}^{\vert n \rangle} =  \sigma_{\it \Theta_{s} \it \Theta_{s}}^{\vert n \rangle}$ for the Fock states $\vert 0 \rangle$, $\vert 1 \rangle$, $\vert 2 \rangle$ and $\vert 3 \rangle$ as functions of $\lambda$, compared with the classical value $\pi^2/12\approx0.82$.}
\end{center}\end{figure}
\begin{figure}\begin{center}
\epsfbox{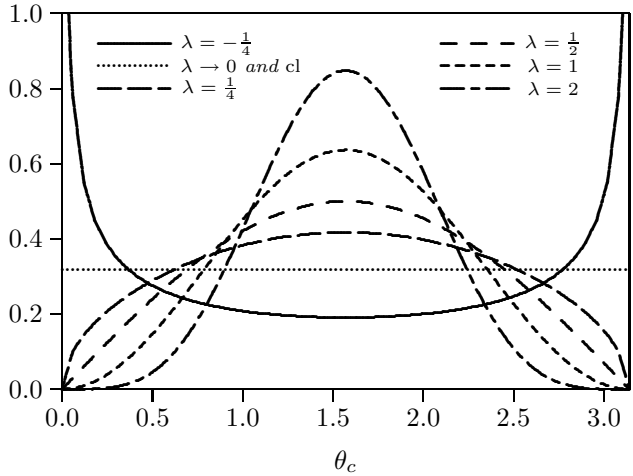}
\caption{\label{F11}Arccosine distributions 
$\mathcal{P}_{\vert 0 \rangle}^{(\Theta_{c})}(\theta_{c})$ for the vacuum state $\vert 0 \rangle$, for $\lambda=-\frac{1}{4}$, $\rightarrow 0$, $\frac{1}{4}$, $\frac{1}{2}$, $1$ and $2$, compared with the classical value $1/\pi \approx 0.32$.}
\end{center}\end{figure}
For polynomials of definite parity 
the expectation values 
$\langle n \vert {\it \hat \Theta_{s}}^{l} \vert n \rangle$ 
vanish for odd powers $l$. In this case  $\langle n \vert {\it \hat \Theta_{c}} \vert n \rangle = \frac{\pi}{2}$ and $\langle n \vert {\it \hat \Theta_{s}} \vert n \rangle = 0$.
The variances squared are shown in Figure~\ref{F7} as functions of the Gegenbauer parameter $\lambda$ for the Fock states $\vert 0 \rangle$, $\vert 1 \rangle$, $\vert 2 \rangle$, $\vert 3 \rangle$. For fixed $\lambda$ and increasing $n$ they approach either from above or from below the classical value $\pi^2/12=0.8225$. The behaviour is similar to the one observed in Figure~\ref{F1} for $\sigma_{CC}^{\vert n \rangle} =  \sigma_{SS}^{\vert n \rangle}$.

The arccosine [arcsine] distributions (\ref{G962}) for a Fock state $\vert n \rangle$ vanish at the $n$ roots of the polynomial $P_{n}(\cos\theta_{c})$ [$P_{n}(\sin\theta_{s})$] and must show therefore an oscillatory behaviour with $n+1$ peaks. At the end points the distributions vanish, stay finite or tend to infinity depending on the behaviour of $\sin \theta_{c}\, w(\cos \theta_{c})$ [$\cos \theta_{s}\, w(\sin \theta_{s})$] for $\theta_{c} \rightarrow 0, \pi$ [$\theta_{s} \rightarrow \pi/2, -\pi/2$].
In the case of the Jacobi polynomials this behaviour is given by $(\theta_{c})^{2\mu+1}$, $(\pi-\theta_{c})^{2\nu+1}$ [$(\frac{\pi}{2} - \theta_{s})^{2\mu+1}$, $(\frac{\pi}{2} + \theta_{s})^{2\nu+1}$].
For the Gegenbauer polynomials ($2\mu+1 = 2\nu+1 = 2\lambda$) the distributions vanish (diverge) at both endpoints for $\lambda > 1$ ($\lambda < 1$) and stay finite for $\lambda \rightarrow 0$ (Chebyshev polynomials of the first kind).
These features can be observed in Figure~\ref{F11} showing the arccosine distributions of the vacuum state $\vert 0 \rangle$ for $\lambda=-\frac{1}{4}$, $\rightarrow 0$, $\frac{1}{4}$, $\frac{1}{2}$, $1$, $2$. 
Here the distribution for $\lambda\rightarrow 0$ (corresponding to the polynomial $T_{0}(x)$) coincides with the classical uniform distribution (\ref{G961}).

The arccosine $\langle \theta_{c} \vert \alpha \rangle$ and arcsine $\langle \theta_{s} \vert \alpha \rangle$ representations of the coherent state (\ref{G400}) can be obtained from (\ref{G800}) by replacing $p_{n}(c)$ and $p_{n}(s)$ by $c_{n}(\theta_{c})$ and $s_{n}(\theta_{s})$, respectively.
The corresponding distributions, 
$\mathcal{P}_{\vert \alpha \rangle}^{(\Theta_{c})}(\theta_{c})$ and 
$\mathcal{P}_{\vert \alpha \rangle}^{(\Theta_{s})}(\theta_{s})$,
are then given by the right-hand-side of (\ref{G840}) with $p_{n}(c)$ replaced by $c_{n}(\theta_{c})$ and ($p_{n}(c),\varphi$) by ($s_{n}(\theta_{s}), \varphi - \frac{\pi}{2}$), respectively.
As a result, the comments made in the paragraph following (\ref{G840}) apply appropriately. Thus, the coherent states with $+ \varphi_{0}$ and $- \varphi_{0}$ ($\varphi_{0}$ and $\pi - \varphi_{0}$) have the same arccosine (arcsine) distributions. Also, the arcsine distributions for $\frac{\pi}{2} \pm \varphi_{0}$ and the arccosine distribution for $\varphi_{0}$ agree if the translation $\theta_{c} = \frac{\pi}{2} - \theta_{s}$ is taken into account.
Neither the trigonometric nor the completely equivalent inverse trigonometric distributions can resolve this phase ambiguity. The $c$ ($s$) distributions `feel' only the values of $\cos\varphi$ ($\sin \varphi$), and the $\theta_{c}$ ($\theta_{s}$) distributions, strictly speaking, only those of $\arccos(\cos \varphi)$ ($\arcsin(\sin \varphi))$. None of the distributions is a true phase-sensitive distribution. Actually, they cannot be one by construction.

\begin{figure}\begin{center}
\epsfbox{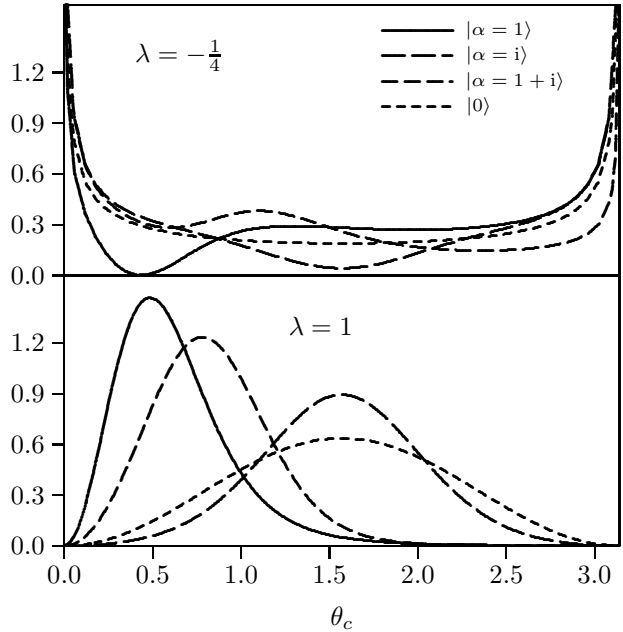}
\caption{\label{F12}Arccosine distributions 
$\mathcal{P}_{\vert \alpha \rangle}^{(\Theta_{c})}(\theta_{c})$ 
for the coherent states with $\alpha = 1$, $\rmi$, $1+\rmi$, $0$, 
and for $\lambda=-\frac{1}{4}$, $1$.
Same distributions for $\alpha = 1$, $-\rmi$, $1-\rmi$, $0$, respectively.}
\end{center}\end{figure}

\begin{figure}\begin{center}
\epsfbox{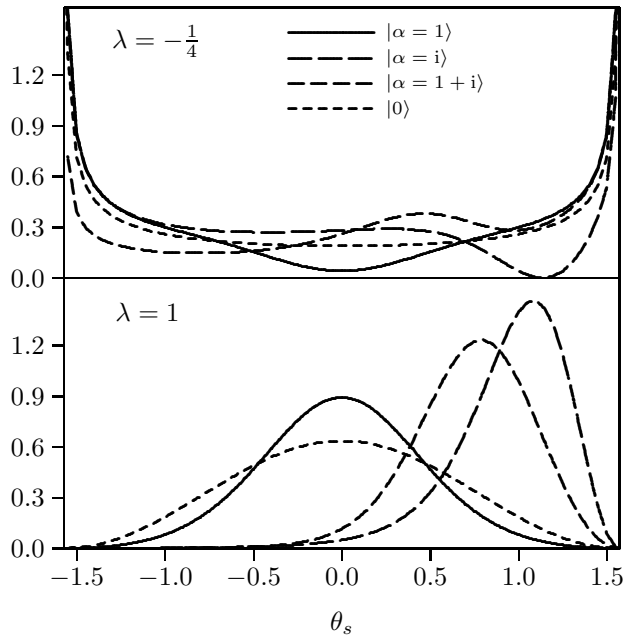}
\caption{\label{F17}Arcsine distributions 
$\mathcal{P}_{\vert \alpha \rangle}^{(\Theta_{s})}(\theta_{s})$ 
for the coherent states with $\alpha = 1$, $\rmi$, $1+\rmi$, $0$, 
and for $\lambda=-\frac{1}{4}$, $1$. 
Same distributions for $\alpha = -1$, $\rmi$, $-1+\rmi$, $0$, respectively.}
\end{center}\end{figure}

We show in Figure~\ref{F12} the arccosine distributions for the coherent states $\vert \alpha\rangle$ with $\alpha = 1$, $\rmi$, $1+\rmi$, $0$ (vacuum state) and for $\lambda=-\frac{1}{4}$, $1$.
The corresponding arcsine distributions are shown in Figure~\ref{F17}. 
The distributions should be compared with the cosine and sine distributions for the same input data in Figures~\ref{F6} and \ref{F16}. 
Here the distributions develop peaks (dips) for $\lambda > 0$ ($\lambda < 0$) too. Their position is close to $\theta_{c} = \varphi$ in Figure~\ref{F12} and to $\theta_{s} = \varphi$ in Figure~\ref{F17}, where $\varphi=0$, $\frac{\pi}{2}$ and $\frac{\pi}{4}$ is the phase of $\alpha$. If we expect a distribution to peak around the value corresponding to the coherent phase, then the distributions with $\lambda < 0$ should be disregarded.

We conclude this section by making two remarks. 
First, we note the identities
\begin{equation}
\cos^2 {\it \hat \Theta_{c}}+\sin^2 {\it \hat\Theta_{c}}=\hat 1 ,\qquad \quad 
\cos^2 {\it \hat \Theta_{s}}+\sin^2 {\it \hat\Theta_{s}}=\hat 1 ,
\end{equation}
showing that in the sense of this trigonometric relation the `right partner' of $\hat C =\cos\it \hat \Theta_{c}$ ($\hat S =\sin \it \hat\Theta_{s}$) is $\sin(\arccos \hat C)$ ($\cos(\arcsin \hat S$)) and not $\hat S$ ($\hat C$).

Second, the Hermitian operators $\it \hat \Theta_{c}$ and $\it \hat \Theta_{s}$ can be used to define two distinct noncommuting unitary operators \cite{cn} by 
\begin{equation}
\hat U_{c}\, =\, \exp(\rmi \it \hat \Theta_{c}) , \qquad \qquad \quad \ 
\hat U_{s}\, =\, \exp(\rmi \it \hat \Theta_{s}) . \label{G980}
\end{equation}
Writing 
$\exp(\rmi \it \hat \Theta ) = \cos \it \hat \Theta + \, \rmi \, \sin \it \hat \Theta$ 
and using (\ref{G94}), we obtain
\begin{eqnarray}
\hat C + \rmi \, \hat S \, &=& \, \hat E \ + (1+\rmi) \hat E_{0} \, = \, 
\frac{1}{2}\, [\, \hat U_{c} + \hat U_{s} + \hat U_{c}^{\dag} - \hat U_{s}^{\dag}\, ] , \nonumber \\
\hat C - \rmi \, \hat S \, &=& \, \hat E^{\dagger} + (1-\rmi) \hat E_{0}\, = \, 
\frac{1}{2}\, [\, \hat U_{c} - \hat U_{s} + \hat U_{c}^{\dag} + \hat U_{s}^{\dag}\, ] .
\label{G985} 
\end{eqnarray}
The combinations $\hat C \pm \rmi \hat S$ yield pure shift operators, as in the Susskind--Glogower and Lerner cases, only for the generalized operators. In the case of the extended operators the displaced combinations ($\hat C - \hat E_{0}) \pm \rmi \, (\hat S - \hat E_{0})$ should be considered instead.

\section{Summary} \label{S6}

In this paper we have introduced cosine and sine operators which generalize (extend) in a specific way the cosine and sine operators of Susskind and Glogower. Our starting point was the observation that the eigenstates of the Susskind--Glogower operators in the Fock basis are given by the Chebyshev polynomials of the second kind. We have extended this relationship to arbitrary polynomials which are orthogonal on the intervall $\mbox{[-1,1]}$ with respect to a weight function. Related with each polynomial set there is a pair of cosine and sine operators determined by the recurrence coefficients for the orthonormal polynomials and with eigenstates given in terms of the polynomials themselves. 
The relationship implies: i) the eigenvalue equations for the cosine and sine operators are satisfied on the basis of the three-term recurrence formula for the polynomials, ii) the orthogonality and iii) the completeness relations for the eigenstates follow directly from the completeness and orthogonality relations, respectively, for the polynomials. Two types of operators emerged naturally, termed generalized or extended operators, depending on wether the weight function is symmetric or not. The distinction can be interpreted physically in terms of normal mode or displaced mode variables. In the case of the classical orthogonal polynomials the Jacobi polynomials with equal (unequal) indices give rise to generalized (extended) operators. The generalized operators belong to the class of operators considered by Lerner and are proper generalizations of the Susskind--Glogower operators. The extended operators define a new class of cosine and sine operators which, to our knowledge, has not been considered so far in this context.

For both types of cosine and sine operators we have introduced lowering and raising exponential operators as well as inverse arccosine and arcsine operators. The eigenstates of the trigonometric and inverse trigonometric operators have the orthonormal polynomials as Fock components, are orthogonal and yield a resolution of unity. This allows them to be used in defining cosine, sine, arccosine and arcsine representations and corresponding probability distributions for arbitrary normalized states. The phases `measured' by the cosine and arccosine distributions on the one side and the sine and arcsine distributions on the other side are in quadrature, as expected. None of the distributions, however, is a true phase distribution.
The still missing eigenstates of the lowering exponential operators of both types are of a different nature and will be considered in a forthcoming paper.

\end{document}